\documentclass{aa} 
\usepackage{graphicx,epsfig}
\usepackage{longtable}
\usepackage{natbib}
\bibpunct{(}{)}{;}{a}{}{,}
\usepackage{txfonts}

\def\vsini{$V\!\sin i$}
\def\vasini{$V_{A}\!\sin i$}
\def\vbsini{$V_{B}\!\sin i$}
\def\teff{T$_{\rm eff}$}

\def\logg{$\log~g$}

\def\kps{km~s$^{-1}$}

\def\theto{$\theta^{\rm 2}$~Tau}
\def\degr{\hbox{$^{\rm o}$}}  

\begin{document}
\title{Spectra disentangling applied to the Hyades binary \theto~AB: new orbit, 
orbital parallax and component properties\thanks{Based on observations obtained 
at the 1.93-m telescope of the {\it Observatoire de Haute Provence}, 
the 1.2-m Mercator telescope at the Roque de los Muchachos Observatory (in the framework of the {\sc Hermes} 
Consortium) and the 1.5-m Wyeth telescope at Oak Ridge Observatory.}}

\titlerunning{Orbit, orbital parallax and masses of \theto~AB}

\author{Torres, K. B. V.$^{1}$, Lampens, P.$^{1}$, Fr\'emat, Y.$^{1}$, 
Hensberge, H.$^{1}$, Lebreton, Y.$^{2,3}$
 \and
\v{S}koda, P.$^{4}$
  }

\authorrunning{Torres, K. et al.}
\offprints{P. Lampens}

\institute{$^{1}$Koninklijke Sterrenwacht van Belgi\"e, Ringlaan 3, B-1180 Brussel, Belgium
\email{kbtorres2003@yahoo.com.br, patricia.lampens@oma.be}\\ 
$^{2}$GEPI \& UMR CNRS 8111, Observatoire de Paris, 92195 Meudon Cedex, France\\
$^{3}$IPR Universit\'e de Rennes 1, Campus de Beaulieu 3, 35042 Rennes Cedex, France\\
$^{4}$ Astronomical Institute, Academy of Sciences of the Czech Republic, 25165 Ond\v{r}ejov\\
}

\date{Received, 2010}

\abstract
{}
{\theto~is a detached and single-lined interferometric-spectroscopic binary as well as the most 
massive binary system of the Hyades cluster. The system revolves in an eccentric orbit with a periodicity 
of 140.7 days. Its light curve furthermore shows a complex pattern of $\delta$ Scuti-type pulsations. 
The secondary has a similar temperature but is less evolved and fainter than the primary. 
In addition, it is rotating more rapidly. Since the composite spectra are heavily blended, the direct 
extraction of radial velocities over the orbit of component B was hitherto unsuccessful. 
Our aim is to reveal the spectrum of the fainter component and its corresponding Doppler shifts in order 
to improve the accuracy of the physical properties of this important "calibrator" system.} 
{Using high-resolution spectroscopic data recently obtained with the {{\sc Elodie}} 
(Observatoire de Haute--Provence, France) and {{\sc Hermes}} (Roque de Los Muchachos, La Palma, Spain) 
spectrographs, and applying a spectra disentangling algorithm to three independent data sets  
including CfA spectra (Oak Ridge Observatory, USA), we derived an improved spectroscopic orbit. We 
next used a code based on simulated annealing and general least-squares minimization to refine the 
orbital solution by performing a combined astrometric-spectroscopic analysis based on the new spectroscopy 
and the long-baseline data from the Mark~III optical interferometer.
} 
{As a result of the performed disentangling, and notwithstanding the high degree of blending, 
the velocity amplitude of the fainter component is obtained in a direct and objective way. 
Major progress based on this new determination includes an improved computation of the orbital parallax 
(still consistent with previous values).
Our mass ratio is in good agreement with the older estimates of Peterson
et al. (1991, 1993), but the mass of the primary is 15--25\% higher than
the more recent estimates by Torres et al. (1997) and Armstrong et al. (2006).
}
{The evolutionary status of both components is re-evaluated in the light of 
the revisited properties of \theto~AB. Due to the strategic position of the components in the turnoff
region of the cluster, the new determinations imply stricter constraints for the age and the
metallicity of the Hyades cluster. We conclude that the location of component~B can be explained 
by current evolutionary models, but 
{the location (and the status) of the more evolved component A is not
trivially explained and requires a detailed abundance analysis of its
disentangled spectrum. The improved accuracy (at the 2\% level) on the stellar masses 
provides a useful basis for the comparison of the observed pulsation frequencies with 
suitable theoretical models.}
}


\maketitle
%

\section{Introduction}

A modern research topic that we are currently pursuing is the study of binary and multiple stars 
with at least one pulsating component. The advantages of studying pulsating components in  
well-detached systems are manyfold, e.g. both the theories of stellar evolution and of stellar 
pulsation can be accurately tested and refined. Accurate component properties compared to suitably 
chosen theoretical isochrones indeed allow to obtain information on the object's age and evolutionary 
path and are usually necessary to help discriminate among various possible pulsation models. 

We selected the $\delta$ Scuti star $\theta^2$ Tau for a careful study for three reasons: (a) 
it is a detached, spectroscopic binary resolved by long-baseline interferometry 
\citep[][hereafter AM06]{2006AJ....131.2643A};
(b) it is a member of the Hyades open cluster at a mean distance of 45~pc \citep{1998A&A...331...81P};
(c) the evolutionary status of the primary component is still under heavy debate 
(\citeauthor{1997ApJ...485..167T} \citeyear[][hereafter TSL97]{1997ApJ...485..167T};
\citeauthor{1999A&A...349..485L} \citeyear[][and AM06]{1999A&A...349..485L}).

$\theta^2$ Tau (HD~28319 = HIP~20894 = 78~Tau) is the most massive main-sequence star of the 
Hyades cluster: it is located at the main-sequence turnoff region of the isochrone best fitting 
the individual members of the cluster \citep[][hereafter LE01]{2001A&A...374..540L}. 
It forms a quadruple system with $\theta^1$ Tau
(= 77 Tau), a common proper motion companion at an angular separation of 5$\farcm$6~. The brighter component 
is a single-lined spectroscopic binary (SB1) which was resolved by long-baseline interferometry 
\citep[]{1991BAAS...23..830S,1992ASPC...32..502P,1992ASPC...32..552H}. The orbital period of the binary is 140.7 
days with an eccentricity of about 0.7~. The primary component, $\theta^2$~Tau~A, has been classified 
as A7~III 
and rotates with \vsini~= 70 \kps~\citep{2006MmSAI..77..174F}.  It is very hard to detect the secondary, 
$\theta^2$~Tau~B, spectroscopically because the 
Doppler shifts are only a fraction of the width of its broad spectral lines. 
The secondary is less evolved and therefore fainter than $\theta^2$~Tau~A. 
TSL97 treated $\theta^2$~Tau as a double-lined spectroscopic binary (SB2) 
using a 2D-cross correlation method in which they considered the extra pull of the secondary in order to obtain 
improved radial velocities of the primary component. However, they were unable to obtain reliable radial velocity 
measurements for $\theta^2$ Tau B and they only observed its (orbital) influence on the velocities of the brighter 
companion. For these reasons, also the mass ratio could not be directly determined. Still, they were able to derive a 
interferometric--spectroscopic orbit and to determine the component masses and the distance of the binary by 
exploring and assuming a range of values for the mass ratio and the rotational velocity of the secondary star. 
The orbital parallax of their solution agrees well with the Hipparcos trigonometric parallax \citep{1997ESASP1200.....P}. 
The outcome is that the components have different projected rotational velocities: TSL97 obtained 
a best fit assuming K$_{B}$ = 38 \kps~and $V_B\!\sin i$ = 110 \kps, with resulting component masses of M$_{A}$ 
= 2.42 $\pm$ 0.30 M$_{\odot}$ and M$_{B}$ = 2.11 $\pm$ 0.17 M$_{\odot}$. Recently, the component masses and luminosities have 
been redetermined by AM06. From their interferometric data coupled to the Hipparcos secular (i.e. proper-motion based)
parallax \citep{2001A&A...367..111D} and a compatible choice of spectroscopic orbits, these authors obtained component masses of M$_{A}$ = 2.15 $\pm$ 
0.12 M$_{\odot}$ and M$_{B}$ = 1.87 $\pm$ 0.11 M$_{\odot}$. 

As a member of a well-studied cluster, both the metallicity and the distance of $\theta^2$ Tau are known within narrow
boundaries. Such a large amount of information concerning the fundamental properties of the components of a detached 
binary system (with components in a different evolutionary phase) 
{ has led} to various attempts of confrontation with 
stellar evolutionary models. Before Hipparcos \citep{1997ESASP1200.....P},
\citet{1992A&A...260..183K} concluded that the evolutionary status of $\theta^2$~Tau was either in the thick H(ydrogen)-shell 
burning phase (without overshooting) or in the H-core burning phase (with overshooting). From the location in a colour-magnitude 
diagram and a best fit with an isochrone of age $\approx$ 630~Myr and metallicity Z = 0.027, TSL97 (using \citet{1994A&AS..106..275B}
's models) concluded that the primary is in a phase near H-core exhaustion, immediately preceding the phase of overall contraction. But, because both binarity and fast rotation may affect the colour indices, the uncertainty remained. 
\citet{1999A&A...349..485L} used the binary to evaluate various stellar evolution models stating that {\it ``the three theoretical models allow to fit 
correctly the system (...)  in agreement with the more recent constraints available about the metallicity of the Hyades cluster''}. 
They found that the primary could be 
in the H-core burning (end) phase or in the H-shell burning (beginning) phase depending 
on its metallicity. In an extensive study of the Hyades cluster, \citet{2001A&A...367..111D} again used $\theta^2$~Tau and concluded 
that the agreement with the CESAM isochrones (including convective core overshoot) was remarkably good. LE01 derived a 
maximum age of 650 Myr and the initial helium content of the cluster by comparing the mass-luminosity relation based 
on a set of five binaries also containing $\theta^2$~Tau with predictions from models appropriate for the Hyades (using Y = 0.26). 
They issued a warning for the stars located in the turnoff region claiming that {\it ``the interpretation is complicated by the 
effects of rotation and overshooting that make either model or photometric data uncertain''} and that {\it ``improvement of the mass 
of $\theta^2$~Tau~A would certainly better constrain the overshooting by anchoring the star more precisely on its isochrone''}. 
More accurate masses is precisely what AM06 obtained. However, the lower masses and/or brighter luminosities do not conform with 
some of the recent stellar evolutionary models, i.e. \citet{2000A&AS..141..371G} and LE01, for the age and metallicity of 
the Hyades. Finally, \citet{2006MNRAS.368.1941Y}
tried to model both components of $\theta^2$~Tau using models including differential 
rotation. They concluded that good agreement was found for $\theta^2$~Tau~A with Z = 0.024 and an age of 700 Myr. However,
the same model could not explain the location of the secondary component in the colour-magnitude diagram. The evolutionary status 
of both components of $\theta^2$~Tau is therefore still a non-trivial issue.

The system has one more attractive feature: its primary component is a typical multiperiodic $\delta$ Scuti star. 
Various multi-site campaigns have been conducted. \citet{1989A&A...214..209B} obtained five closely spaced 
and stable frequencies, all of which had amplitudes below 0.01 mag. They discarded rotational splitting since it 
could not explain the observed frequency separations and proposed a mixture of modes of different {\it l} and {\it m} 
values. \citet{1996A&A...313..571K} discussed a large set of radial velocity and line profile data from which
up to seven frequencies emerged with only three frequencies in common with the previous analysis. They suggested long-term 
($>$ 6 yr) amplitude variability and a combination of low and high degree modes. Amplitude variability on a 10 yr 
time scale is also claimed by \citet{1997PASP..109..217L}. Both components are located within the lower Cepheid 
instability strip (where the $\delta$ Scuti stars are observed) but it seems well established that the more massive 
primary is the pulsating star \citep{1989A&A...214..209B,1996A&A...313..571K}. The results of a vast multi-site campaign \citep{2002MNRAS.336..249B}
revealed the presence of 11 frequencies in the range 10$-$15 c/d and 2 frequencies in the range 26$-$27 c/d 
\citep[see also][]{2002A&A...382..157P}. After having modelled the light-time corrections, they attributed the first 
11 frequencies to pulsations of the primary component while the last two frequencies probably originate from the secondary 
component. Currently, because of the limited frequency range of the detected modes, the complex frequency spectrum does 
not allow a proper mode identification.

Because they are located in the turnoff region of the cluster, the exact locations of both stars in the 
H-R diagram allow to distinguish between different evolutionary models and different isochrones, thus making {them}
extremely useful for the constraints they impose on the chemical composition and the age of the Hyades (provided 
that their physical properties can be determined very accurately). As  ``calibrator stars'', they could indicate 
how important the mixing processes (such as convective core overshooting or rotational mixing) are in the internal 
layers
of stars from the Hyades cluster (LE01). The knowledge of accurate fundamental component properties furthermore 
holds good potential for a reliable pulsation modelling of the two stars.   

The outline of this article is the following: the observations are
presented in Sect.~2 while the analysis method of spectra disentangling
is explained in some detail in Sect.~3. In Sect.~4, we will compare
the component spectra obtained from the application of the spectra disentangling 
method with an observed spectrum of a reference star as well as with synthetic spectra. Especially in the
case of the fainter component, not easily discernable in the observed
composite spectra, this will show that the mathematically reconstructed spectra 
are plausible from the viewpoint of physics. Sect.~5 deals with
the orbital analysis obtained by combining both spectroscopic and interferometric
data. In Sect.~6, we thoroughly discuss the quality of our results and
their implication in the light of the component's evolutionary stages.    
We end by mentioning the new perspectives and the planned future work.

\section{Observations}\label{observ}

The spectroscopic data consist of (a) 44 \'echelle spectra
(R=42000; S/N\footnote{From hereon, S/N = S/N per resolution element}$\sim$200$-$300 in V; 
resolution element of 7.1 \kps) obtained by us with {\sc Elodie} 
at the 1.93 m telescope of the Observatoire de Haute--Provence covering
the wavelength range from 389.5 to 681.5 nm; 
(b) 13 \'echelle spectra (R$\sim$85000; S/N$\sim$220$-$340 in V; 
resolution element of 3.5 \kps)
obtained by us with the {\sc Hermes} spectrograph at the 1.2 m Mercator telescope
located at the Roque de Los Muchachos Observatory, La Palma, covering the wavelength
range from 377 to 900 nm; 
(c) 70 \'echelle spectra (R$\sim$35000; S/N$\sim$50; resolution element of 8.5 \kps)
obtained by TSL97 using the Center for Astrophysics (CfA) spectrograph
mounted on the 1.5 m Wyeth reflector at the Oak Ridge Observatory, covering a wavelength 
range of 26~$\AA$ around 519~nm;
(d) 16 single-order coud\'e spectra (R$\sim$25000; S/N$\sim$100-200; resolution element 
of 14 \kps) obtained at the 2-m telescope of the Ond\v{r}ejov Observatory 
of the Astronomical Institute of the Academy of Sciences
of the Czech Republic, covering the wavelength range 517--589~nm.

The {\sc Elodie} data were reduced using the {\sc Intertacos} pipeline \citep{1996A&AS..119..373B} 
available at the telescope,
while the {\sc Hermes} data were treated using the {\sc Hermes} reduction pipeline 
\citep{2010A&A..00..000V}. These reduction 
procedures perform the order extraction, the offset and flat--field correction, and the 
wavelength calibrations. The resulting wavelength scale was corrected for the Earth's motion
relatively to the barycenter of the solar system using the {\sc Iraf} software package. 
{The spectra collected at the Ond\v{r}ejov Observatory were reduced with the
procedure described by \citet{2002PAICz..90...22S}.}

In total, 127 spectra\footnote{ The 44 spectra
from the {\sc Elodie} spectrograph can be retrieved directly from the {\sc Elodie} database
(http://atlas.obs-hp.fr/elodie/). The 70 spectra from the Harvard Smithsonian Center
for Astrophysics (CfA) and the 13 ones from the {\sc Hermes} spectrograph
are available upon request from the authors of the paper Torres et al. 1997
and from us, respectively.} covering the entire
orbital cycle and 
{full range in radial velocity amplitude}
were available
for the analysis\footnote{Unfortunately, the Ond\v{r}ejov
single-order coud\'e spectra did not enter the subsequent 
disentangling procedure due to their lower resolution and 
lower S/N (caused by unfavourable weather conditions
during the observations) compared to the sets of \'echelle spectra.}.
We furthermore made use of 34 best-fit angular separations ($\rho$) and
position angles ($\theta$) derived from the measurements obtained by
\citet{1992hrii.conf..673A}, \citet{1992ASPC...32..552H} and AM06 with the Mark~III long-baseline optical interferometer.

\section{Spectra disentangling}\label{specdis}
\begin{table}
\caption{Spectral intervals used in the spectra disentangling analysis.}
\centering
\begin{tabular}{ccccc}
\hline\hline\noalign{\smallskip}
Index        & $\lambda_i$        &  $\lambda_f$       & N & $\frac{\ell_B}{\ell_A}$ \\ [-3.5pt]
             &  {(\AA)}       & {(\AA)}        &   &                         \\
\hline\noalign{\smallskip}
R1 & 4660.37 & 4725.97 & 2363  & 0.364    \\ 
R2 & 4759.53 & 4961.50 & 4520  & 0.364    \\ 
R3 & 5178.81 & 5292.47 & 3700  & 0.360    \\ 
R4 & 5421.21 & 5475.42 & 1700  & 0.359    \\ 
R5 & 5520.36 & 5610.08 & 2700  & 0.359    \\ 
R6 & 6090.67 & 6127.51 & 1035  & 0.355    \\ 
R7 & 6127.55 & 6187.31 & 1665  & 0.353    \\   
\hline
\end{tabular}
\tablefoot{$\lambda_i$ and $\lambda_f$ refers to the initial and final wavelength in \AA ngstr{\"o}ms; N refers to
the number of bins contained in the observed composite spectra;
and ${\ell_B}/{\ell_A}$ corresponds to the light ratio of the components determined as
explained in Sect.~\ref{specdis}.}
\label{tab:reg}
\end{table}
We used the spectra disentangling code {\sc FDBinary} V.3
(Rel.~30.01.09) developed by \citet{2004ASPC..318..111I}, which 
determines the individual contributions of the components to the composite spectra 
together with the orbital parameters in a self-consistent way. The 
{\sc FDBinary} runs are based on the input of observed spectra at epochs which, 
in an ideal case, uniformly cover the orbital 
radial velocity range. The observed spectrum is assumed to be the combination 
of two time-independent component spectra that are shifted in wavelength with respect
to each other (according to the derived Keplerian orbit). 
The code uses the multi-dimensional, non-linear optimization technique of the 
(downhill-)simplex \citep{1992nrfa.book.....P} to determine the orbital
parameters, whereas the intrinsic component spectra are computed by the  
algorithm of spectra separation (using the radial velocities estimated from 
Kepler's equations).  
Commonly, many runs starting from various points  
in a subspace of the orbital parameter space are launched,
and different sizes for the initial simplex can be explored. Convergence is attained when the size 
of the simplex shrinks below a specified level in a specified maximum 
number of iterations. The separation algorithm works on the 
Fourier components of the observed spectra using singular value decomposition 
for quasi-singular sets of equations.

Our analysis was divided in two steps: (a) the determination
of the orbital parameters from a selected wavelength region
and (b) the computation of the intrinsic component spectra,
keeping the orbital parameters fixed during the convergence,
for other wavelength regions (cf. Table~\ref{tab:reg}).

\begin{figure}[t]
\center
\includegraphics[width=9cm,clip=]{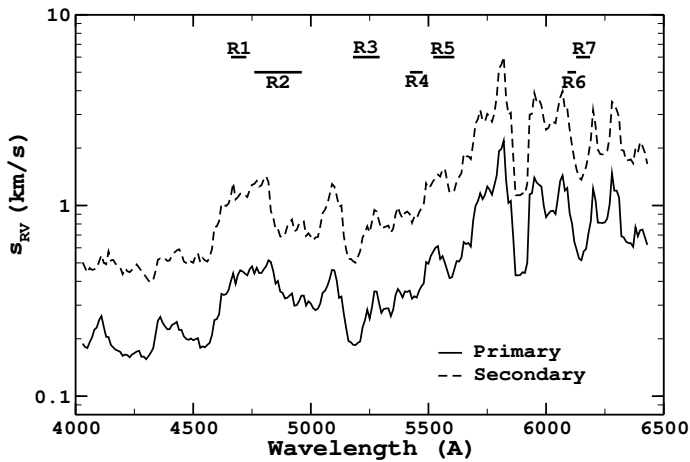}
\caption{Uncertainty of the radial velocity information content as a function of wavelength, 
assuming S/N=100. The standard deviation, S$_{RV}$, was computed over a 60~\AA~wide interval, 
shifted with a step of 10~\AA. The primary and secondary are represented, respectively
by a full and a dashed line. Each wavelength region (Table~\ref{tab:reg}) is represented by 
a small horizontal line. }
\label{fig:rms}
\end{figure}

\begin{figure}[t]
\center
\includegraphics[width=9cm,clip=]{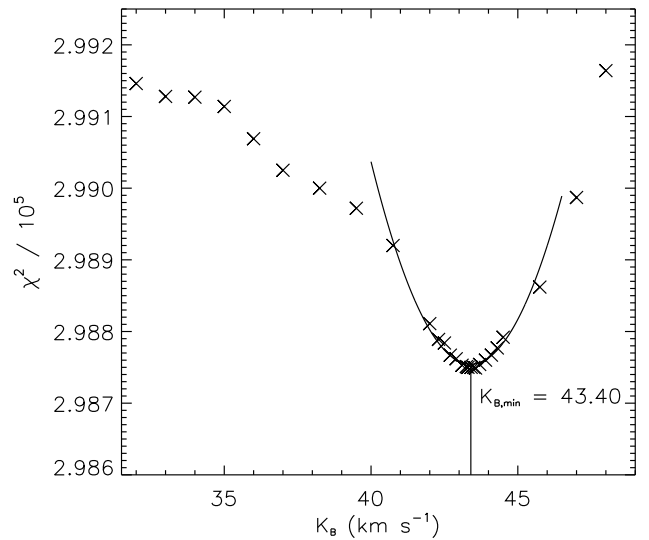}
\caption{Distribution of the $\chi^2$ of the solutions derived by spectra disentangling as a function of 
the radial velocity semi-amplitude of the secondary.}
\label{fig:kb}
\end{figure}

In step (a), the search for the orbital parameters 
was done using 127 spectra weighted according to their S/N ratio in 
the wavelength region R3 (cf. Table~\ref{tab:reg}). This region was chosen 
for it allows the highest accuracy on the determination of the component 
radial velocities among all selected regions 
{(cf. Fig.\ \ref{fig:rms}: the standard deviation shows a deep 
local minimum in that region) }.
Since the CfA spectra cover the large range of radial velocity amplitude near 
periastron passage but have a smaller wavelength range of 26~{\AA}, we only used 
a part of R3. Moreover, the CfA, {\sc Elodie} and {\sc Hermes} spectra together 
provide sufficient spectral coverage in orbital phase.
The basis for this limitation to the range of the CfA spectra is the experience 
that, in a very eccentric orbit, the coverage of the total velocity range is more 
important than the extension of the spectral range (nonetheless, a significantly
broader spectral range would help if it also included a minimum of spectra 
near periastron passage). 
The orbital period was fixed to the accurate value determined by TSL97 (Table~\ref{tab:elements}). 
The {\sc Elodie} and {\sc Hermes} spectra were oversampled to the velocity resolution of the CfA spectra.
The code minimizes the squared differences between the observed composite spectra and the model spectra
computed with respect to a chosen set of orbital elements. The objective function, $\chi^2$, can thus be 
written as follows:\\
\begin{equation}
\chi^2 = \sum_j W(t) [I_{obs,t}(j) - (\ell_{A} I_A(j+\delta_{A,t})+\ell_{B} I_B(j+\delta_{B,t})]^2,
\label{Eq1}
\end{equation}
where $I_{obs,t}(j)$ represents a set of normalized composite spectra at orbital phase $t$ 
and pixel $j$ with weight $W(t)$, $I_A(j)$ and $I_B(j)$ are the two component spectra, $\delta_{A,t}$ 
and $\delta_{B,t}$~ are the component Doppler shifts. The component's respective (time-independent) 
light contributions are indicated by $\ell_{A}$ and $\ell_{B}$.  
Since a complex structure with secondary minima was revealed in the $\chi^2$ space (due to the broad lines 
of the secondary component), a search for the global minimum was executed for a grid of $K_B$-values 
(Fig.~\ref{fig:kb}). The final spectroscopic orbit computation was performed with $K_B = 43.4$ \kps~determined 
from this grid search (cf. Table~\ref{tab:elements}, upper panel). We adopted an uncertainty of the order 
of 0.5 \kps~on this value, corresponding to treating as equivalent all the solutions with 298840 $<$ $\chi^2$ $<$ 
298850, as indicated by Fig. ~\ref{fig:kb}. Unfortunately, Fourier disentangling does not (yet) include any error 
estimation of the orbital elements. 
In step (b), the component spectra were reconstructed applying the separation algorithm with this 
spectroscopic orbit in other wavelength regions (Table~\ref{tab:reg}).

\begin{figure}[t]
\center
\includegraphics[width=9cm,clip=]{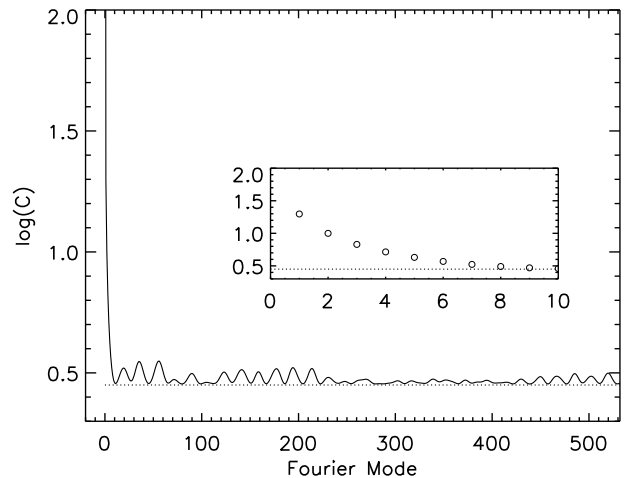}
\caption{Logarithm of the condition number ($\log$ C) versus the Fourier mode
computed for \theto~using the radial velocities derived with FDBinary.   
}
\label{fig:cn}
\end{figure}

In order to verify whether the numerical problem is well-conditioned,
the condition numbers $C_m$ (i.e. the ratio of the largest to the 
smallest eigenvalue in the covariance matrix) for the set of equations for each Fourier mode $m$ {was} 
computed (see Fig.~\ref{fig:cn}). 
{$\log C$ is a rough measure of the loss of
precision on the Fourier amplitudes in the component spectra, expressed in units of number of digits.
One digit is lost, relative to the bottom value of $\log C$, in the
$m = 1$ mode  and this provoked a low-level  sinusoidal undulation in the
reconstructed spectra (see \citeauthor{2008A&A...482.1031H} 
\citeyear{2008A&A...482.1031H} for more explanation). }
{Here,} mode 0 is completely undetermined ($\log C = \infty$)  
due to the lack of eclipses (dilution of spectral lines not varying with 
time).   
{The} low-level sinusoidal undulation in the 
reconstructed spectra due to the lack of constraint on the luminosity ratio 
was removed self-consistently 
(\citeauthor{2004ASPC..318..111I} \citeyear{2004ASPC..318..111I} and 
\citeauthor{2008A&A...482.1031H} \citeyear{2008A&A...482.1031H})
This corresponds to replacing the best purely 
mathematical solution by the most acceptable solution from the viewpoint 
of physics (flat continuum, conserving observed line blocking) at the 
expense of a completely insignificant increase in the $\chi^2$-value. 

The indeterminacy of the $m = 0$ mode implies that the light ratio 
between the two components has to be estimated from external information.
Therefore, it was determined
for each spectral region using the differential magnitude ($\Delta m$) measurements
reported by AM06 (cf. their Table 3). 
A linear dependence of $\Delta m$ on $\lambda$ represents these measurements within their uncertainties.
A weighted fit gives:
\begin{eqnarray} 
\Delta m & = & 1.131 + 0.020~(\lambda - 6361)/1000 \label{dm:equ}.  \\
 & \pm & 0.012 \pm 0.011 \nonumber 
\end{eqnarray} 
The zero-point at the weighted average wavelength of 6361~\AA~ was chosen 
such that the constant and the gradient are uncorrelated.
Eq.\ (\ref{dm:equ}) fixes the light ratio (see black line in Fig.\ \ref{fig:lum}),
used to renormalize the component spectra,
with the central wavelength $\lambda = \lambda_{{\mathrm c}_j}$
of each region $j$ (cf. Table~\ref{tab:reg}) substituted in the equation.

\begin{figure}[t]
\center
\includegraphics[width=9cm,clip=]{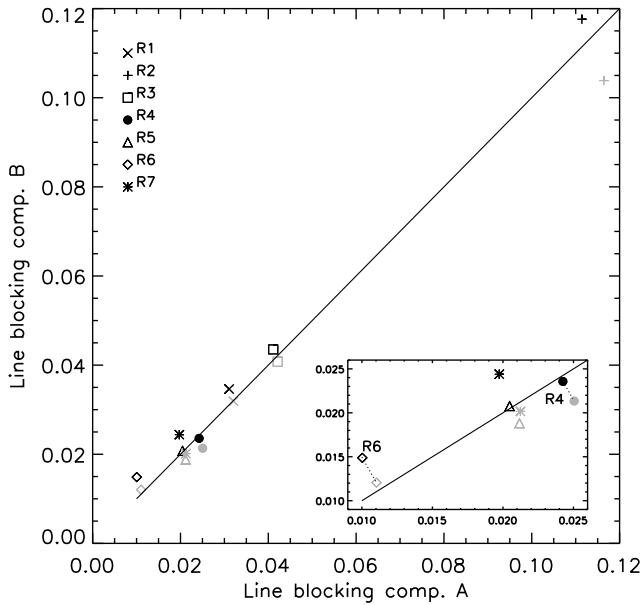}
\caption{Line blocking of \theto~AB for each of the 7 spectral regions
described in Table~\ref{tab:reg}. See Sect.~\ref{specdis}.}
\label{fig:lb}
\end{figure}

\begin{figure}[t]
\center
\includegraphics[width=9cm,clip=]{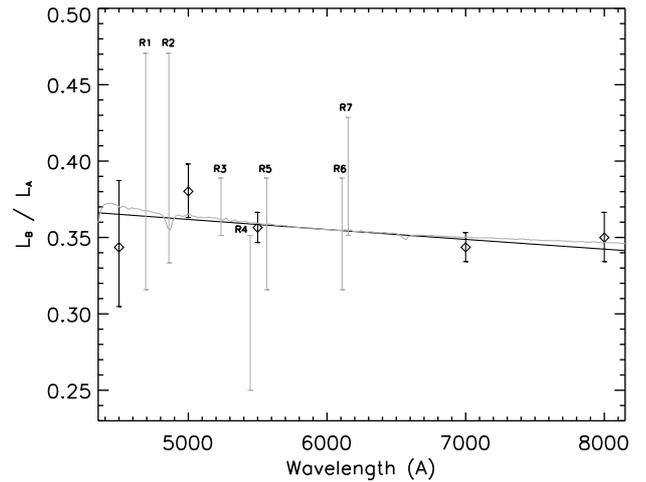}
\caption{ 
Determination of the light contributions of \theto~AB,
using the disentangled component spectra.
The grey vertical bars represent the acceptable values 
for $\ell_B/\ell_A$, for each spectral region (Table~\ref{tab:reg}).
The $\ell_B/\ell_A$ predicted by AM06 are also shown (i.e. black diamonds with error bars),
as well as Eq.\ (\ref{dm:equ}) (straight black line). 
{The grey line represents the slope of the flux ratio curve 
of two synthetic spectra with a difference of temperature corresponding to +200~K,
in the sense (component~B - component~A).
See Sects.~\ref{specdis} and \ref{compspecs1}.}
}
\label{fig:lum}
\end{figure}

In a first step, we fixed the value of the lu\-mi\-no\-si\-ty ratio in order to verify whether
the disentangled component spectra are consistent with the assumption of equal line blocking,
since both components have very similar colours \citep{1993AJ....105.2260P} 
and must have the same overall chemical composition.
Fig.~\ref{fig:lb} compares the line blocking for each spectral region in both components.
The result is sensitive to sub-percent changes in the continuum levels, in
other words, to the solution of the Fourier mode $m$ = 0. Physical constraints on
pseudo-continuum data points deliver limits to the coupled continuum
placement (represented by grey and black symbols). The diagonal in the
figure represents the locus of equal line blocking. 
As seen from this figure, the assumption of equal line blocking is nowhere (i.e. in none
of the studied regions) in contradiction with the allowed range of the choice of the
continuum levels. Under this assumption, the offset in continuum level of the primary's
spectrum may be larger than the previously derived limits (see, for example, R6 
in Fig.~\ref{fig:lb}), but it will never exceed the level of
0.5\%. Such an offset remains insignificant in the context of the present study. 

Therefore, in a second step, we {evaluated which range of continuum positions 
was acceptable around the one for assumed equal line blocking, and, in turn,
estimated the range of possible monochromatic luminosity ratios from the disentangled 
component spectra.} 
Fig.~\ref{fig:lum} shows that {our} spectroscopic estimates are in concordance
with the interferometric ones, {though} are less accurate than the latter.
The grey intervals represent the possible va\-lues for $\ell_B/\ell_A$
derived from the renormalization procedure \citep{2004ASPC..318..111I} used to correct the disentangled
component spectra into physically meaningful spectra.
\section{Component Spectra}\label{compspec}

\subsection{Comparison with a reference stellar spectrum}\label{compspecs1}

We compared the reconstructed component spectra to the observed 
spectrum of the $\delta$ Scuti star HD~2628 
(\vsini~$\sim$20 \kps) which is classified as A7~III, as is also the case of 
the binary system \theto~AB (AM06). 
For visual concordance, the reference spectrum was shifted
by 39.5 \kps~and broadened by 66 \kps~and 118 \kps~in order to match
the spectra of component A and B respectively.

\begin{figure*}
\center
\includegraphics[width=17cm,clip=]{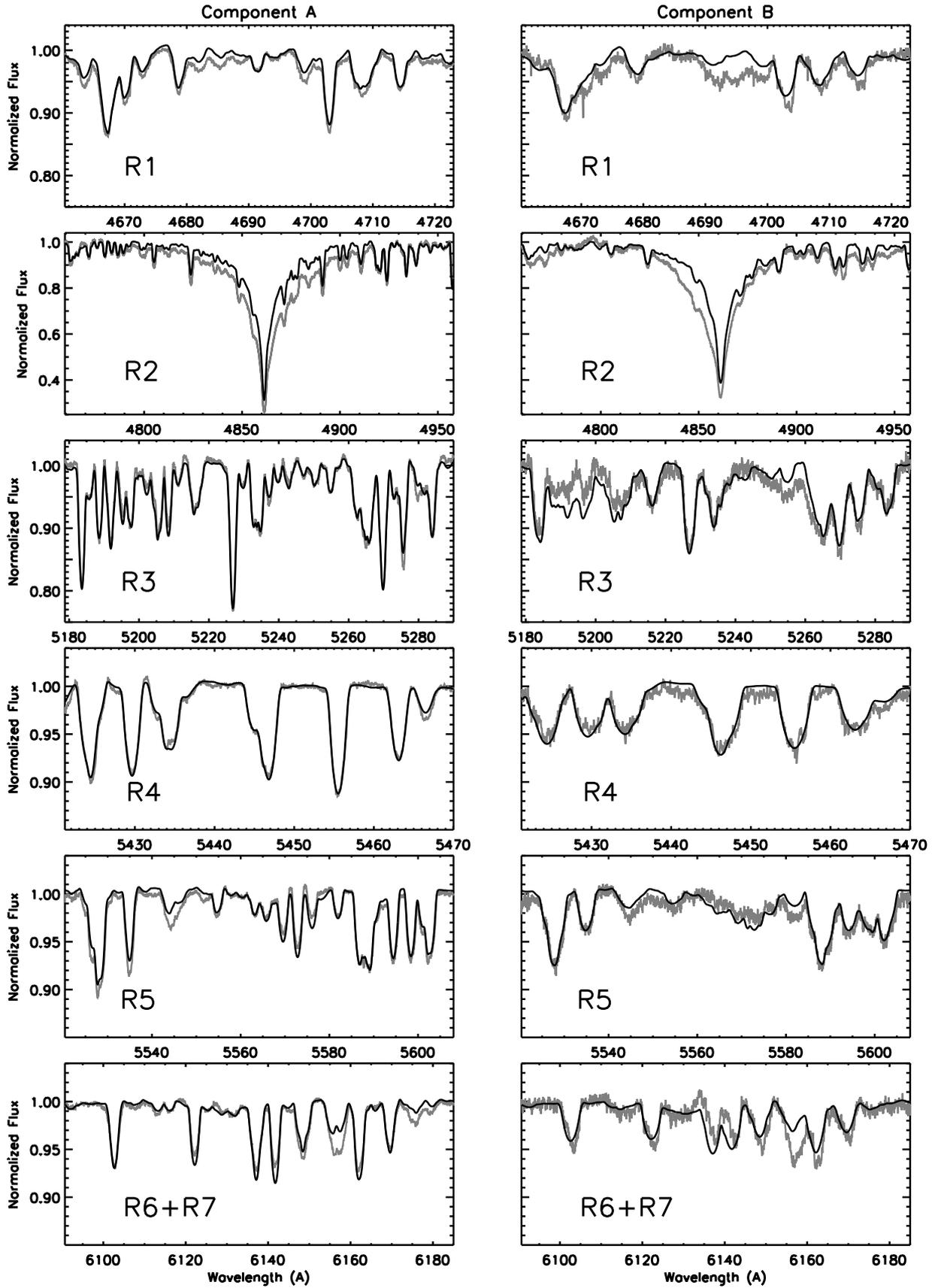}
\caption{Comparison between the reconstructed component spectra of
\theto~AB (grey) and the spectrum of HD~2628 (black), for the regions listed in
Table~\ref{tab:reg}.
The latter spectrum was broadened by 66 and 118 \kps, respectively
for components A and B, and shifted by 39.5 \kps. 
}
\label{fig:ref}
\end{figure*}

Fig.~\ref{fig:ref} shows how well the reconstructed component spectra 
agree with the reference star spectrum, especially for   
the spectrum of component A which matches better the luminosity class of  
HD~2628. The spectra of component B may, at the one per cent level, still be 
affected by low-amplitude wiggles corresponding to larger uncertainties in 
the solution of the $m = 2$ or even slightly higher Fourier modes.
The position and shape of the lines coincide well, showing  
the power of the 
{spectra} disentangling technique in 
reconstructing component 
spectra without any a priori assumptions about their spectral features.
Note, however, that the shape of the Balmer H$\beta$--line {(Fig.~\ref{fig:ref}, panels R2)}
indicates a slight temperature difference between the two components of
\theto~and the reference star {(HD~2628 is about 400~K cooler
than \theto~AB)}.
\subsection{Comparison with synthetic spectra}\label{comsynt}

In order to determine \vsini~and the system velocity 
$\gamma$, we also compared the reconstructed spectra with synthetic spectra  
in the regions R3 to R5 of Table~\ref{tab:reg} and averaged the 
results. 

Independently fitting the H$\alpha$ and H$\beta$--lines of 
each component spectrum, an average effective temperature
over these two H--lines was determined: $7800\pm170$~K for component~A and B. 
We also derived two limits for the temperature difference, $\Delta{\rm T_{eff,B-A}}$, by computing
synthetic spectra with slightly different temperatures and by imposing that their flux ratio
curve should pass through all interferometric measurements (cf. black error bars in Fig.~\ref{fig:lum}).
Note that this curve is more sensitive in the bluest part of the wavelength range. These limits suggest 
a temperature difference, $\Delta{\rm T_{eff,B-A}}$, ranging from +100 to +500~K (best fit at +200~K, cf. 
grey line), 
which is also consistent with our derived values of the component temperatures, considering
their uncertainty. 

A small difference in effective temperature is also confirmed by the interferometric measurement of the colour 
difference of $\sim-0.006$ mag \citep{1993AJ....105.2260P}. For both components, \logg~was estimated 
from the averaged effective temperatures and the components luminosities (see Sect.~\ref{finorb}).
Table~\ref{tab:ele2} summarizes the atmospheric parameters of \theto~ using the
synthetic spectra and compares them with the parameters adopted (except \vbsini~and 
${\rm \gamma}$, which were determined by TSL97).

\begin{figure*}
\center
\includegraphics[width=18cm,clip=]{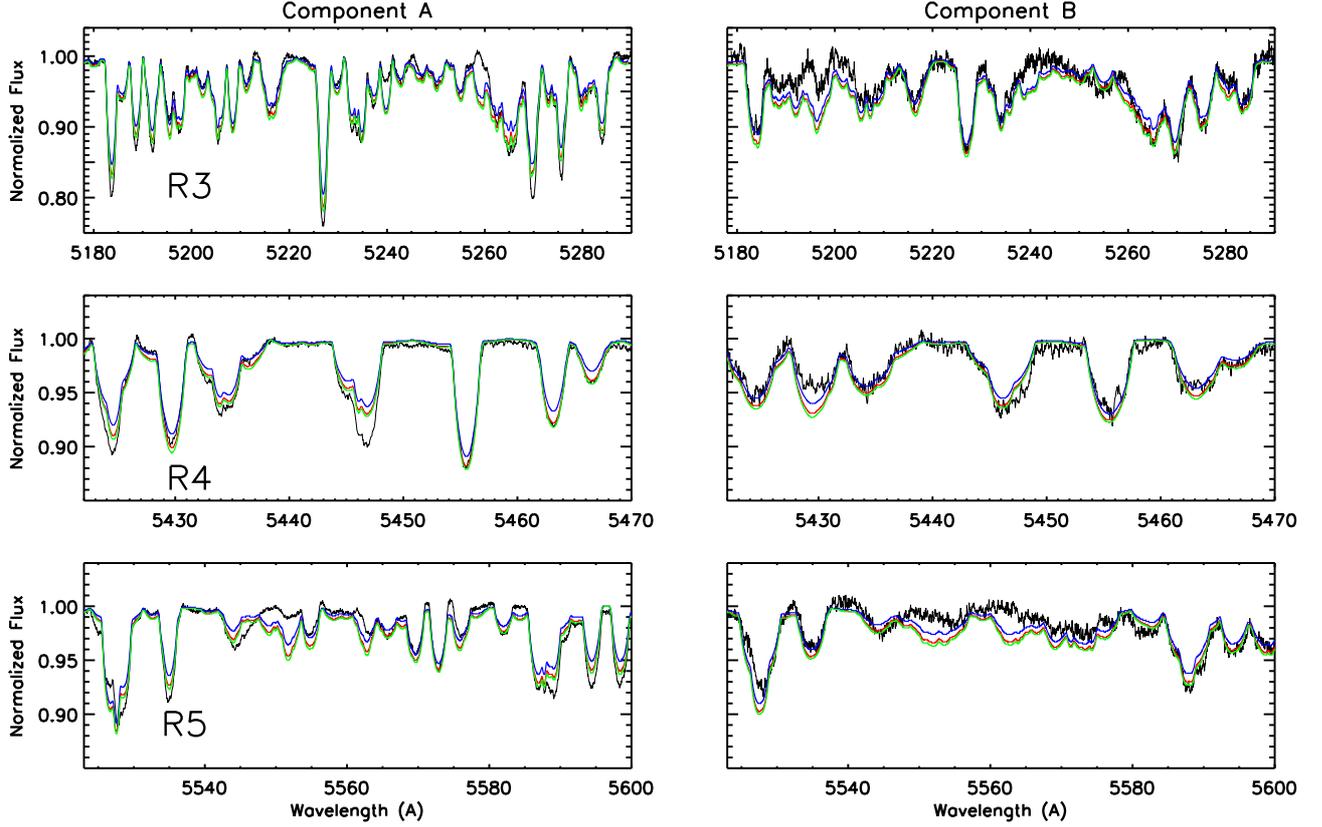}
\caption{ 
Intrinsic spectra of \theto~A and B (in black) compared to synthetic
spectra computed for the Hyades metallicity ([Fe/H]=+0.14 in red; [Fe/H]=+0.19 in green;
LE01), and the solar metallicity (in blue).
The luminosity ratio is fixed to 0.35. The \teff,
\vsini~and \logg~values were fixed to those ones described in the Table~\ref{tab:ele2}.
} 
\label{fig:metal}
\end{figure*}

The Str${\rm \ddot{o}}$mgren photometric data provided by the General
Catalogue of Photometric Data (\citeauthor{1997A&AS..124..349M} \citeyear{1997A&AS..124..349M}) for 
\theto~: V = 3.41, b-y  = 0.100, $m_{1}$= 0.197, $c_{1}$= 1.012, u-b  = 1.606,
and $\beta$  = 2.831, confirm these effective temperatures. Indeed, a combined effective temperature 
${\rm T_{eff,AB}}$ = 7928~K results from an updated version of the standard calibration by \citet{1985MNRAS.217..305M}
(Napiwotski, private communication).
\begin{table}[h]
\begin{center}
\caption{Atmospheric parameters of \theto~A and B.
}
\begin{tabular}{lcc}
\hline\hline\noalign{\smallskip}
Atmospheric& Synthetic    & TSL97 \\
parameters &  spectra     &  \\
\noalign{\smallskip}\hline\noalign{\smallskip}
$V_A\!\sin i$ (\kps)   & 68.4 $\pm$ 1.5  &  70\tablefootmark{2}  \\
$V_B\!\sin i$ (\kps)   & 113  $\pm$ 6    &  110 $\pm$ 4\\
${\rm T_{eff,A}}$ (K) & 7800 $\pm$ 170\tablefootmark{1}        &  8250\tablefootmark{2} \\
${\rm T_{eff,B}}$ (K) & 7800 $\pm$ 170\tablefootmark{1}        &  --- \\
$\log~g_{A}$     & 3.6  $\pm$ 0.1 &   4.0\tablefootmark{2}     \\
$\log~g_{B}$     & 3.9  $\pm$ 0.1 &   4.5\tablefootmark{2}     \\
${\rm \gamma}$ (\kps) & 39.3 $\pm$ 0.9  &  39.5 $\pm$ 0.2\\
\noalign{\smallskip}\hline
\end{tabular}
\tablefoot{The atmospheric parameters and their standard deviations 
were computed from the component spectra using synthetic spectra 
(second column), in this work. The values determined by \citet{1997ApJ...485..167T} 
are also listed for comparison (third column).\\
\tablefoottext{1}{average of two values obtained using H$\alpha$ and H$\beta$--lines;}
\tablefoottext{2}{adopted by TSL97.}
}
\label{tab:ele2}
\end{center} 
\end{table}
Fig.~\ref{fig:metal} shows the disentangled component spectra and the
synthetic spectra computed for the solar metallicity and for two possible 
values of the Hyades metallicity proposed by LE01
([Fe/H]=+0.14 and [Fe/H]=+0.19). 
The observed line strengths suggest that higher metallicities might be better 
than solar ones, as expected for two stars belonging to the Hyades cluster, but 
this claim should be confirmed by a detailed abundance analysis. 
\section{Combined Orbital Analysis}\label{com_ana}

Accurate orbital solutions of stellar systems are obtained when
different techniques are used together to provide input data to a single least-squares 
analysis, thereby leading to the simultaneous determination of all the parameters, 
also including an independent determination of the distance to the system.     
To this aim, a combined interferometric-spectroscopic analysis
of \theto~was performed \citep{2009AIPC.1170..446L}.
Unfortunately, to our knowledge, no software is available that 
directly combines the orbital solution of spectral 
disentangling with the astrometry. Thus, radial velocities must be 
specified, with uncertainties. 
These uncertainties will define the relative weights of the radial velocities
with respect to those of the relative positions (interferometry).
We used 2 {\rm x} 127 radial velocities obtained from the {\sc Elodie}, {\sc Hermes} and CfA 
spectra and 34 best-fit angular separations $\rho$ and position angles $\theta$ 
(Sect.~\ref{observ}).  

\subsection{Input radial velocities and their uncertainties}\label{input}

There exist two options for the input radial velocities:  
(a) we may derive radial velocities applying the cross-correlation 
technique using the reconstructed component spectra as templates, or 
(b) we may compute radial velocities from the spectroscopic orbit derived with 
{\sc FDBinary}. The first delivers ``observed'' (noisy) measurements, 
but the orbit derived from them will not exactly correspond to the 
orbit derived with {\sc FDBinary}. The orbit that would be derived 
from the radial velocities resulting from the cross-correlation is 
of slightly inferior quality than the self-consistent orbit obtained 
from {\sc FDBinary}. 
Hence, option (b) was preferred. Nevertheless, we checked that choosing 
option (a) 
{is consistent with} the combined astrometric--spectroscopic 
solution.  
We also evaluated the radial velocity information content in the 
spectral region, assuming that random noise largely dominates the 
error budget in order to derive associated uncertainties (see e.g. Fig.\ \ref{fig:rms}). 
In this sense, these estimates are lower limits to the true uncertainties. 
The principles for a single-star spectrum are elaborated in several papers, 
e.g. \citet{1999A&AS..136..591V} and for a binary in \citet{2000A&A...358..553H}. 
The latter uncertainties essentially contain a correction factor (relative to the 
single-star case) for the correlation at a given orbital phase between the spectral 
gradients in the Doppler-shifted component spectra, and a multiplicative factor  
inversely proportional to the contribution of the component to the total 
light \citep{2010A&A..000..000V}.
Typical uncertainties thus derived are given by the following median values:
$\sigma({\rm RV}_A)$ = 0.15 ~and $\sigma({\rm RV}_B)$ = 0.68 \kps~({\sc Elodie} spectra);
$\sigma({\rm RV}_A)$ = 0.10 ~and $\sigma({\rm RV}_B)$ = 0.54 \kps~({\sc Hermes} spectra);
$\sigma({\rm RV}_A)$ = 0.65 ~and $\sigma({\rm RV}_B)$ = 3.10 \kps~(CfA spectra).
In addition, we also made sure that these uncertainties in radial velocity 
are compatible with the scatter that one would obtain by (re)computing the radial 
velocities from the range of acceptable orbits derived with FDBinary which correspond 
to equivalent solutions in the $\chi^2$ plane (see Fig.~\ref{fig:kb}). Table~\ref{tab:rv:asp}
lists the uncertainties on the component radial velocities derived with FDBinary.

\subsection{Final orbital elements}\label{finorb}

The combined solution was computed using the code {\sc vbsb2}, which performs 
a global exploration
of the parameter space followed by a local least-squares minimization 
\citep{1998A&AS..131..377P}.
The essence of the method resides in a simultaneous adjustment of various data types
performed in two steps: (a) the minimum of the objective function is globally searched 
following the principles of simulated annealing \citep[SA,][]{1953JChPh..21.1087M} and (b) the 
best estimate of a large number of trials provides the starting point 
of a local least-squares minimization (using the procedure of Broyden-Fletcher-Goldstrab-Shanno).

Table~\ref{tab:elements} (bottom panel) lists the orbital elements 
and their standard deviations obtained after 200 runs with SA, exploring a small 
interval around the period provided by TSL97. 
The orbital elements were slightly updated in the combined
astrometric--spectroscopic solution 
(cf. Table~\ref{tab:elements}), without significant effect on the derived
component spectra. Interestingly, our determinations of the mass ratio (0.754 $\pm$ 0.001) and K$_{B}$ are in close agreement with
the first estimates derived by \citet{1991ASPC...13..592P} and \citet{1993AJ....105.2260P} (see Table~1 in AM06 for the 
mass ratio estimate using Peterson's corrected radial velocities).
\begin{table}
\begin{center}
\caption{Orbital solutions of \theto~A and B from various techniques.} 
\tabcolsep 3.0pt
\begin{tabular}{lcc}
\hline\hline\noalign{\smallskip}
\multicolumn{3}{c}{Spectroscopic Orbital Solution ({\sc FDBinary})} \\
\noalign{\smallskip}\hline\noalign{\smallskip}
\multicolumn{1}{l}{Parameter} & \multicolumn{1}{c}{This work} \\
\noalign{\smallskip}\hline\noalign{\smallskip}
P (days)               & 140.72816\tablefootmark{1}  \\
T                      & 1993.0751  \\
e                      & 0.734662   \\
$\omega_B$ ($^{\circ}$)& 235.88     \\
K$_{A}$ (\kps)         & 32.74      \\
K$_{B}$ (\kps)         & 43.40\tablefootmark{2} ($\pm$ 0.5, cf. text) \\
\noalign{\smallskip}\hline\noalign{\smallskip}    
\multicolumn{3}{c}{Interferometric Orbital Solution (AM06) } \\
\noalign{\smallskip}\hline\noalign{\smallskip}
P (days)             &  140.72816\tablefootmark{1} \\
T                    &  1989.60701 $\pm$ 0.00003 \\
e                    &   0.73725  $\pm$ 0.00036 \\
a (mas)              &  18.796 $\pm$ 0.056 \\
i ($^{\circ}$)       &  47.61  $\pm$ 0.09 \\
$\Omega$ ($^{\circ}$)& 173.73  $\pm$ 0.07 \\
$\omega_A$ ($^{\circ}$)& 55.40  $\pm$ 0.06  \\
{\it $\pi_{sec}$ (mas)} & {\it  22.30\tablefootmark{3}} {\it $\pm$ 0.36} \\
\noalign{\smallskip}\hline\noalign{\smallskip}    
\multicolumn{3}{c}{Astrometric-Spectroscopic Orbital Solution} \\
\noalign{\smallskip}\hline\noalign{\smallskip}
   Parameter        & This work        & TSL97  \\
\noalign{\smallskip}\hline\noalign{\smallskip}
P (days)             &  140.7302  $\pm$ 0.0002  &  140.72816 $\pm$ 0.00093 \\
T                    & 1995.00144 $\pm$ 0.00002 & 1993.0752  $\pm$ 0.0008 \\
e                    & 0.7360     $\pm$ 0.0003  & 0.7266  $\pm$ 0.0049  \\
a (mas)              & 18.91    $\pm$ 0.06  & 18.60\tablefootmark{4} $\pm$ 0.20 \\
i ($^{\circ}$)       &  47.8  $\pm$ 0.1  &  46.2\tablefootmark{4}  $\pm$ 1.0 \\
$\Omega$ ($^{\circ}$)& 353.82  $\pm$ 0.09  & 171.2\tablefootmark{4}  $\pm$ 1.8 \\
$\omega_B$ ($^{\circ}$)& 235.41  $\pm$ 0.08  & 236.4  $\pm$ 1.1 \\
$q=\frac{M_{\rm B}}{M_{\rm A}}$ & 0.754  $\pm$ 0.001 & 0.873 $\pm$ 0.048\\
$\pi_{orb}$ (mas)    & 20.90   $\pm$ 0.10  & 21.22  $\pm$ 0.76 \\
K$_{A}$ (\kps)  & 32.95   $\pm$ 0.04  &  33.18 $\pm$ 0.49 \\
K$_{B}$ (\kps)            & 43.68   $\pm$ 0.14  &  38\tablefootmark{5} $\pm$ 2 \\
M$_{A}$ ($M_{\odot}$) & 2.86    $\pm$ 0.03  &  2.42  $\pm$ 0.30 \\
M$_{B}$ ($M_{\odot}$) & 2.16    $\pm$ 0.02  &  2.11  $\pm$ 0.17 \\
rss (all)    & 1.23e+3    &   8.18e+03 \\ 
System mass ($M_{\odot}$)& 5.02  $\pm$ 0.09   &  4.54 $\pm$ 0.51 \\
Time span (yr)    &   20.0  &  6.3  \\
\noalign{\smallskip}\hline
\end{tabular}
\tablefoot{
The standard deviations including orbital parallax and masses are shown in the bottom panel.\\ 
\tablefoottext{1}{Fixed (adopted from TSL97)}
\tablefoottext{2}{Fixed (see Sect.~\ref{specdis})}
\tablefoottext{3}{Adopted from \citet{2001A&A...367..111D} (not an orbital element)}
\tablefoottext{4}{ Adopted from PS92}
\tablefoottext{5}{Fixed (after a search in 2-D space).}
}
\label{tab:elements}
\end{center}
\end{table}
{Table~\ref{tab:rv:asp} lists the radial velocities corresponding to the orbit
derived with FDBinary, their uncertainties (cf. Sect.~\ref{input}) and the differences
in the sense (Combined solution - FDBinary orbit).
}
The combined astrometric--spectroscopic orbit predicts radial velocities that differ less than 0.1 (for comp.~A) 
or 0.14 \kps~(for comp.~B) from the pure spectroscopic orbit derived with FDBinary
over the whole time interval covered by the observations and over all orbital phases,
except for a narrow phase interval around periastron where the differences grow to 0.7 and 
0.9 \kps~respectively. This corresponds to a change of $\sim1$ \% in the mass of the
components.
\begin{table}[h]
\begin{center}
\caption{\theto~AB radial velocities.}
\tabcolsep 2.5pt
\tiny{
\begin{tabular}{ccr@{}r@{}rcr@{}r@{}rcr}
\hline\hline\noalign{\smallskip}
HJD $+$       & Orbital & \multicolumn{3}{c}{${\rm RV}_A$}& {(C $-$ } & \multicolumn{3}{c}{${\rm RV}_B$} & { (C $-$ }& Set  \\ 
  2400000     & phase   & \multicolumn{3}{c}{(\kps)}      &   {FDB) }      & \multicolumn{3}{c}{(\kps)}       &  {FDB) }      &      \\ 
\hline \\ [-6pt]
47844.67396 & 0.68250& { $+$3.217  }& $\pm$ &  1.208&  $+$0.034 & { $-$4.264   }& $\pm$ &   5.456&  $-$0.038 & 1\\
47942.61109 & 0.37842& { $-$9.362  }& $\pm$ &  1.130&  $+$0.022 & { $+$12.409  }& $\pm$ &   5.235&  $-$0.021 & 1\\
47957.58553 & 0.48483& { $-$5.451  }& $\pm$ &  0.747&  $+$0.035 & { $+$7.225   }& $\pm$ &   3.442&  $-$0.038 & 1\\
48142.89616 & 0.80161& { $+$11.265 }& $\pm$ &  0.773&  $+$0.001 & { $-$14.933  }& $\pm$ &   3.593&  $+$0.005 & 1\\
48145.81918 & 0.82238& { $+$13.202 }& $\pm$ &  0.701&  $-$0.012 & { $-$17.498  }& $\pm$ &   3.262&  $+$0.023 & 1\\
48158.82114 & 0.91477& { $+$26.806 }& $\pm$ &  0.697&  $-$0.185 & { $-$35.530  }& $\pm$ &   3.274&  $+$0.251 & 1\\
\multicolumn{11}{l}{\ldots}\\
\noalign{\smallskip}\hline
\end{tabular}
\tablefoot{The radial velocities (RV) were derived relatively to the center of mass 
($\gamma = 39.3$ \kps) with the code {\sc FDBinary}. (C-FDB) refers
to the differences in the sense {\bf C}ombined astrometric--spectroscopic solution - {\bf FDB}inary orbit.
The uncertainties were computed as explained in Sect.~5.1. 
The Heliocentric Julian Day (HJD) and orbital phase corresponding to each radial velocity are also included.
Set 1: CfA data; Set 2: {\sc Elodie} data; Set 3: {\sc Hermes} data. A complete version
of this table is available on-line. }
\label{tab:rv:asp}
}
\end{center}
\end{table}
The visual (interferometric) orbit recently derived by AM06 comprises 7 orbital elements of 
which one, namely the orbital period, was adopted from TSL97. The remaining six elements (T, 
a, e, i, $\Omega$, $\omega$) are in very good agreement: the difference between AM06 and the
new combined solution amounts to 0.4$\sigma$ for T, 1.4-$\sigma$ for a, 2.7-$\sigma$ for e, 
1.4-$\sigma$ for i, 0.4-$\sigma$ for $\omega$, and 1.2$\sigma$ for the node $\Omega$.  Note 
that the latter element is modified by 180$\degr$ with respect to our solution 
as we follow the spectroscopic convention (while TSL97 followed \citet{1992ASPC...32..502P}, hereafter PS92). 
The marginally significant change in the value of the eccentricity 
places it now in-between the discordant values of TSL97 and AM06.
As an additional check, we verified that
the residuals in both position angle and angular separation show no systematic offset and the
standard deviations, $\sigma_{\mathrm (O-C)_\theta} = 1.2\degr$ and
$\sigma_{\mathrm(O-C)_\rho} = 0.47$ mas,
agree with the published error bars on the measurements. 

In conclusion, the astrometric part of the combined solution is in excellent agreement 
with the pure astrometric solution computed by AM06, except (marginally) for the eccentricity. 
Our value is also constrained by the spectroscopic data, and lies in-between the value derived
by AM06 and that of TSL97 (see Table~\ref{tab:elements}). Also note that the uncertainties 
of the visual orbital elements are very similar to those published by AM06. 

The new orbital parallax has a relative error, $\sigma_\pi$/$\pi$, of 0.5\%, which is several times more 
precise than the one of TSL97 ($\pi$ = 21.22 mas, $\sigma_\pi$/$\pi$ = 3.6\%) and the ones measured by the 
Hipparcos astrometric satellite (\citeauthor{1997ESASP1200.....P} \citeyear{1997ESASP1200.....P}: $\pi$ = 21.89 mas, 
$\sigma_\pi$/$\pi$ = 3.8\%; \citeauthor{2007A&A...474..653V} \citeyear{2007A&A...474..653V}: $\pi$ = 21.69 mas,
$\sigma_\pi$/$\pi$ = 2.1\%). 
Though the value of the new parallax is slightly smaller, it remains compatible with these formerly derived values, within 
the larger uncertainties of these previous determinations. However, the new value is in disagreement 
with the secular parallax derived by \citet{2001A&A...367..111D} (see Table~\ref{tab:elements} ($\sigma_\pi$/$\pi$ 
= 1.6\%) and Sect.~\ref{stat1}).

\begin{figure}[t]
\vspace{0.5cm}
\centering
\includegraphics[height=6cm, clip=]{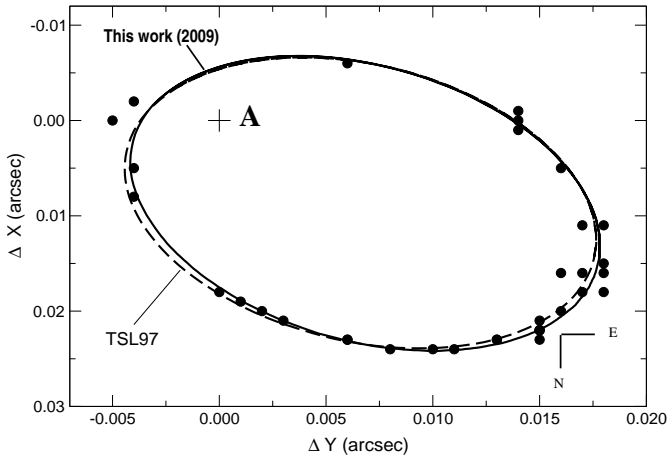}
\caption{Combined orbital solution plotted with the astrometric data 
from \citet{2006AJ....131.2643A}.}
\label{fig:1}
\end{figure}

\begin{figure}[h]
\vspace{0.5cm}
\centering
\includegraphics[height=6cm, clip=]{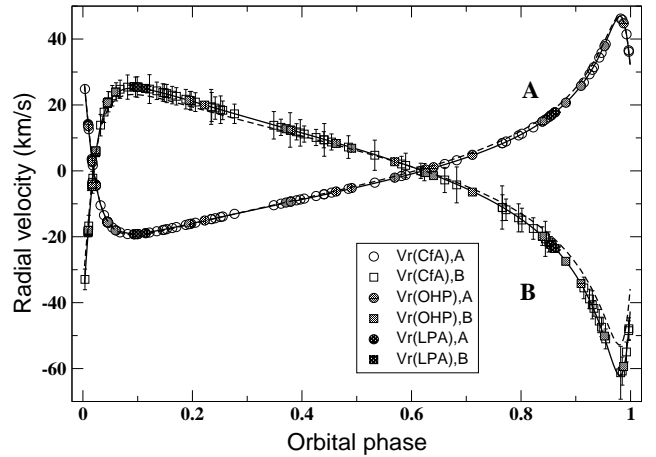}
\caption{Combined orbital solution (full lines) plotted with component radial 
velocities { (symbols ``$\circ$'' for component A and ``$\square$'' for component B) derived 
with {\sc FDBinary}.} Combined orbital solution  derived by TSL97 (dashed lines). 
The different data sets obtained with the {\sc Elodie} (Observatoire de Haute--Provence, OHP),
{\sc Hermes} (Roque de Los Muchachos Observatory, LPA) and CfA spectrographs 
are represented by { the different shades of grey,} see legend.}
\label{fig:2}
\end{figure}

The combined orbital solution is graphically illustrated by Figs.~\ref{fig:1} 
(astrometric observations and combined solution) and~\ref{fig:2} 
(spectroscopic observations and combined solution). The astrometric--spectroscopic orbital 
solution previously derived by TSL97 is shown for comparison. We remark that these authors 
adopted some visual orbital elements of the preliminary orbit published by PS92 
in their combined analysis.  The most conspicuous difference between both studies (ours -- TSL97) 
is the larger radial velocity amplitude of component B, while there is excellent agreement for 
the radial velocity curve of component A over the entire phase range.
Note that, even though a homogeneous coverage in orbital phase was achieved, we still do not have 
a perfect coverage in radial velocity in the critical part of the orbit where the radial velocity is changing 
fastest (only 13\% of the spectra have a radial velocity larger than 24.8 \kps~for component~A).  

At this stage, it is relevant to compare and explain the uncertainties quoted in Table~\ref{tab:elements}. 
{ Here}, we list the formal errors of a (local) least-squares minimization process. These show a 
significant improvement of (at least) a full order of magnitude over the errors listed by TSL97. However, 
TSL97 did {\it not} compute their solution using full least-squares analysis (since the interferometric data 
were not available to them). In addition, some of their uncertainties were derived by way of Monte Carlo simulations 
assuming Gaussian errors on the various parameters. 
As already mentioned, with regard to the astrometric part of the solution, TSL97 adopted the visual orbital elements
and the corresponding errors from PS92. 
Compared to this and other previous analyses 
performed with (subsets of) the Mark~III data, the astrometric orbit derived by AM06 shows a major improvement in quality: 
from their Table~3, a factor of about 10 was gained in the accuracy of the visual orbital elements over 
that of the previous studies. Our work merely confirms the very high quality of their orbital solution. 
 
On the other hand, with regard to the uncertainties in the spectroscopic part of the combined solution, 
we notice that the higher quality (higher S/N) of the 44 {\sc Elodie} and 13 {\sc Hermes} spectra  
{ has led} to 
the extraction of 5-10 times more precise radial velocities: the quality of the radial velocities is expected 
to be 5 times better if the ratio of their typical uncertainties is considered (e.g. {\sc Elodie} versus CfA, 
see Sect.~\ref{input}), and 10 times better if the ratio of the typical rms residual is considered (TSL97 derived 
a rms residual of 1.7 \kps in RV$_{A}$ whereas we have rms residuals of 0.15 and 0.20 \kps in RV$_{A}$ and RV$_{B}$, 
respectively, over the entire data set). The gain might be somewhat larger considering the larger amount of spectra 
(127 instead of 70 spectra), as well as the robustness of the applied disentangling technique (the radial velocities 
in the observed spectra are bound by a Keplerian orbit which probably lowers the bias in the different parameters) 
and the improved astrometric orbit. But the somewhat less favourable distribution over the orbital velocity range of 
the new spectra in comparison with the CfA ones, 
{ might have limited the gain achieved.}
Hence, an overall improvement of a 
factor of (at least) 5 can be expected compared to TSL97. As a matter of fact, Table~\ref{tab:elements} suggests a gain 
with a factor slightly larger than 10 in the radial-velocity amplitudes. 

This difference can be understood as these formal errors remain underestimates of the true errors. For example, the error mentioned for K$_{A}$ is $\pm$ 0.04 \kps, 
which is about 3 times better than the median error in radial velocity extracted for that component (see Sect.~\ref{input}). This 
seems reasonable. However, the error mentioned for K$_{B}$ is $\pm$ 0.14 \kps, which is about 5 times better than the median error 
in radial velocity extracted for that component (see Sect.~\ref{input}). An uncertainty 2 (perhaps 3) times as large for K$_{B}$ might 
be closer to the true uncertainty value, in particular if we also consider the shape of the $\chi^2$ minimum in Fig.~\ref{fig:kb}. Indeed, 
the formal error based on an increase by a unit of $\chi^2$ is 0.15 km/s (assuming a perfect parabolic shape near minimum), i.e. 
similar to the uncertainty listed in Table~\ref{tab:elements}. However, such an estimate corresponds to a strict lower limit of the 
uncertainty since it assumes random errors. With resampled data and the ambiguity in tracing of the continuum levels, uncertainties 
on subsequent pixels get correlated, which unavoidably contributes to introducing some bias in one spectrum with respect to the others. 
To stay on the safe side, we will adopt from hereon an error twice as large for K$_{B}$, i.e. $\pm$ 0.28~\kps~(instead
of $\pm$ 0.14~\kps). As a consequence of the laws of error propagation, we will also consider a larger error 
contribution of the semi-axis major $\alpha_B$ (the semi-axis major of component B with respect to the centre of 
mass expressed in \kps) and on M$_A$ by a factor of 2. The former leads to an error budget on the orbital parallax 
increased by a factor of 1.4 (determined by the ratio of the apparent and the true semi-axis major).  
In Table~\ref{tab:fund-properties}, we summarize all redetermined component properties including
the increased error budget and we compare the new values with those of, respectively, AM06 and TSL97. 
With respect to AM06, the present total mass is 20\% larger, most of which is due to the difference 
in parallax value. With respect to TSL97, the total mass is 10\% larger, with half of the increase 
due to the updated parallax and the other half due to the increased sum of the radial velocity amplitudes.

Adopting the new parallax of Table~\ref{tab:elements} and $\Delta m_V$ = 1.11 $\pm$ 0.01 mag 
(derived from Eq.\ (\ref{dm:equ})) 
we obtain the following component absolute magnitudes: M{\sc v}$_{A}$ = 0.33 $\pm$ 0.03 mag 
and M{\sc v}$_{B}$ = 1.44 $\pm$ 0.03 mag. This determination is more accurate than before 
(thanks to the improved parallax determination) and remains in good agreement with the one 
provided by TSL97. 
However, it differs from that by AM06 
at the 3$\sigma$-level. 
\begin{table}[h]
\begin{center}
\caption{Fundamental 
properties of \theto~A and B components.}
\begin{tabular}{l@{}c@{}c@{}c}
\hline\hline\noalign{\smallskip}
  Component A     &  This work        &   AM06    &  TSL97  \\
\noalign{\smallskip}\hline\noalign{\smallskip}
${\rm T_{eff,A}}$ (K)      & 7800 $\pm$ 170  & --- & 8250\tablefootmark{1} \\
\vasini~(\kps)            & 68.4 $\pm$ 1.5  & --- & 70\tablefootmark{1}  \\
${\rm log(L/L_{\odot})}$   & 1.77\tablefootmark{6} $\pm$ 0.05 & 1.74 $\pm$ 0.05 & 1.78 $\pm$ 0.07 \\
M$_{A}$ ($M_{\odot}$)      & 2.86 $\pm$ 0.06 & 2.15 $\pm$ 0.12 & 2.42 $\pm$ 0.30 \\
M$_{V}$ (mag)              & 0.33 $\pm$ 0.03 & 0.48 $\pm$ 0.05 & 0.37 $\pm$ 0.08 \\
\logg~(cm~s$^{-2}$)        & 3.6 $\pm$ 0.1  & --- & 4.0\tablefootmark{1} \\
\noalign{\smallskip}\hline\noalign{\smallskip}
  Component B     &  This work        &   AM06    &  TSL97  \\
\noalign{\smallskip}\hline\noalign{\smallskip}
${\rm T_{eff,B}}$ (K)      & 7800 $\pm$ 170 & --- & 8250\tablefootmark{1} \\
\vbsini~(\kps)             & 113  $\pm$ 6 & --- &  110 $\pm$ 4 \\
${\rm log(L/L_{\odot})}$   & 1.32\tablefootmark{6} $\pm$ 0.04 & 1.29 $\pm$ 0.06 & 1.34 $\pm$ 0.07 \\
M$_{B}$ ($M_{\odot}$)      & 2.16 $\pm$ 0.02 & 1.87 $\pm$ 0.11 & 2.11 $\pm$ 0.17 \\
M$_{V}$ (mag)              & 1.44 $\pm$ 0.03 & 1.61 $\pm$ 0.06 & 1.47 $\pm$ 0.08 \\
\logg~(cm~s$^{-2}$)        & 3.9  $\pm$ 0.1  & --- & 4.5\tablefootmark{1} \\
\noalign{\smallskip}\hline\noalign{\smallskip}
                  &  This work        &   AM06    &  TSL97  \\
\noalign{\smallskip}\hline\noalign{\smallskip}
$q=\frac{M_{\rm B}}{M_{\rm A}}$ & 0.754 $\pm$ 0.002 & 0.873\tablefootmark{3} & 0.873 $\pm$ 0.048 \\
$\Delta m_{550nm}$ (mag)    & 1.11 $\pm$ 0.01\tablefootmark{4}  & 1.12 $\pm$ 0.03 & 1.10 $\pm$ 0.01\tablefootmark{5}\\
$\pi_{orb}$ (mas)          & 20.90 $\pm$ 0.14 & 22.30 $\pm$ 0.36\tablefootmark{2} & 21.22 $\pm$ 0.76 \\
$\sigma_{\pi_{orb}}$ (\%)  & 0.7  &  1.6 &  3.6 \\
Tot. mass ($M_{\odot}$)  & 5.02  $\pm$ 0.12 &  4.03 $\pm$ 0.20 &  4.53 $\pm$ 0.51 \\
$\sigma_{Sum}$ (\%)&  2  &  5   &  11\\
Method used & VB-SB orbit & VB orbit + $\pi_{dyn}$ & VB-SB orbit \\
\noalign{\smallskip}\hline
\end{tabular}
\label{tab:fund-properties}
\end{center}
\tablefoot{The fundamental properties of \theto~A and B derived in this work (second column), by 
\citet{2006AJ....131.2643A} (third column) and by \citet{1997ApJ...485..167T} (fourth column) 
are described here. The listed uncertainties
may be slightly different from the formal uncertainties in Table~\ref{tab:elements} as they take into account
the increased error budget of K$_B$.\\
\tablefoottext{1}{ adopted;}
\tablefoottext{2}{\citet{2001A&A...367..111D};}
\tablefoottext{3}{from \citet{1997ApJ...485..167T};}
\tablefoottext{4}{from Eq.\ (\ref{dm:equ})}
\tablefoottext{5}{from \citet{1993AJ....105.2260P}}
\tablefoottext{6}{M$_{Bol,\odot}$ = 4.75 mag}
}
\end{table} 
\begin{figure}[ht]
\vspace{0.5cm}
\centering
\includegraphics[height=6cm,clip=]{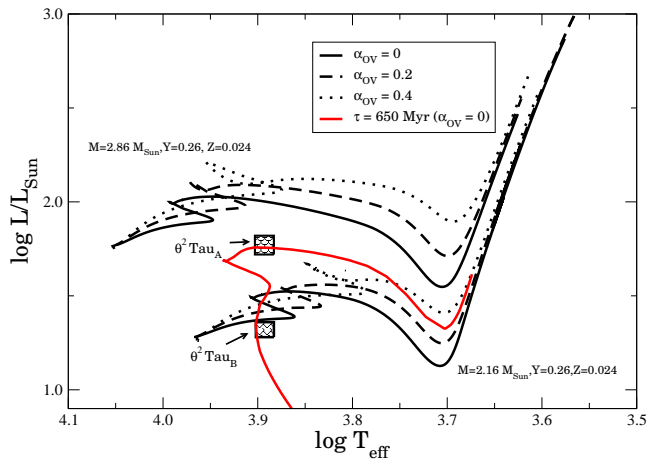}
\caption{
Location of \theto~AB in the Hertzsprung--Russell diagram. 
The curves represent evolutionary tracks for the masses corresponding to \theto~AB, the metallicity of the Hyades 
(following LE01) and three levels of overshooting, $\alpha_{\rm{OV}}$, (different black curves as indicated in the legend) and 
an isochrone of age of 650 Myr (red curve), all computed using the code {\sc Cesam} \citep{2008Ap&SS.316...61M}.}  
\label{fig:evo}
\end{figure}
\section{Comparison with evolutionary models}
\subsection{Previous Status}
\label{stat1}

Various attempts to compare the components' locations in the Hertzsprung-Russell (H-R) diagram with 
evolutionary models have already been done, though not always with great success. The importance
of finding the correct evolutionary models may also have implications for the entire Hyades cluster. 
\citet{1999A&A...349..485L} tested 3 independent sets of stellar evolutionary tracks: those of the Geneva and Padova groups 
(cf. \citet{1999A&A...349..485L} for the references) and those of \citet{1992A&AS...96..255C}.
They found some discrepant results from isochrone fits for 3 visual binaries using the Hipparcos 
parallaxes (including \theto) and were unable to find a single isochrone passing through 
the locations of both components of V818~Tau (van Bueren 22). However, LE01 remarked that the models previously used 
(taken from the literature) did not match the cluster's chemical composition. Using the 
location of V818~Tau, these authors were able to exclude the isochrone at 625~Myr 
with the solar-scaled abundance for He (Y=0.28), and derived the improved value Y = 0.255 $\pm$ 0.009. 
They furthermore concluded that an isochrone with ([Fe/H]=+0.19, Y=0.27) matched best the observed 
main-sequence of the Hyades, whereas a chemical composition with both ([Fe/H]=+0.19, Y=0.27) and 
([Fe/H]=+0.14, Y=0.26) fitted well 
the locations of the 5 considered binaries (including \theto). In the turnoff region, rotational 
effects complicate the interpretation but the overall agreement is good (see Fig.~12 from LE01). 
However, in order to discriminate between the different models 
(e.g. with or without overshooting),
the accuracy on the component masses, also of \theto, was still insufficient.
 
AM06 adopted the dynamical parallaxes based on the Hipparcos proper motions \citep{2001A&A...367..111D}
to revise the orbital solution of \theto~(we recall that this parallax is inconsistent with our 
determination). These authors derived a total mass of 4.03 $\pm$ 0.20 M$_{\odot}$.
Using the spectroscopic mass ratio of TSL97, they computed the component masses and luminosities. They were
however unsuccessful at modelling the component locations in the mass-luminosity and the Hertzsprung-Russell 
diagrams, even though they corrected the luminosities and the colours for the rotational effects assuming 
3 different inclination angles: in each case, the masses appeared too small for the observed luminosities. 
Therefore, they claimed that the component properties of \theto~did not match the current evolutionary tracks. 
However, the He abundance used in their model fits is not appropriate for the Hyades.

\subsection{Current Status}

The revised component properties (cf. Table~\ref{tab:fund-properties}) and, maybe more importantly, 
the higher accuracy with which they were obtained, make it worthwhile to review the location 
of the components of \theto~in the evolutionary diagram. An isochrone model with the age and the 
chemical composition of the Hyades should be fitting the parameter box of both stars within the 
quoted uncertainties. We may also expect that sharper constraints might be put on the age and the abundance 
determination of the cluster, given the fact that the accuracies are (expected to be) of the order of 2\% 
on the component masses and of the order of 4\% on the component luminosities (i.e. an improvement
with a factor of 6-7 with respect to previous determinations). 
This was already the case with V818~Tau (LE01), an eclipsing-spectroscopic binary whose component masses
are known within 1\% \citep{1988ApJ...333..256P}.  

In Fig.~\ref{fig:evo}, we plotted the locations of \theto~A and B in the H-R diagram using the temperatures
and the masses from Table~\ref{tab:fund-properties}. We compare these locations with theoretical 
evolutionary tracks adopting the Hyades composition derived by LE01 ([Fe/H] = +0.14 and Y = 0.26).
Three different values of overshooting were considered: zero overshoot (full lines), and overshoot
values of 0.2 and 0.4, in units of the pressure scale height $H_p$ 
(dashed, respectively dotted lines). As can be clearly seen from Fig.~\ref{fig:evo},
the track with $M$ = 2.16 $M_{\odot}$ and zero overshoot passes through the box of component~B (at an
age of $\approx$ 600~Myr), while the tracks with $M$ = 2.86 $M_{\odot}$ are way too luminous compared 
to the location of the box representing component~A. Nevertheless, an isochrone model of age 650~Myr 
and with the same chemical composition (thin line) passes through both boxes. We conclude that component~B 
is indeed on the main sequence (as previous authors did), whereas the interpretation for component~A remains 
enigmatic: neither convective core overshooting nor rapid rotation can be invoked to explain the observed 
discrepancy. Taking into account gravitational darkening due to rapid rotation
would shift its observed location in the wrong direction (its non-rotating counterpart would be hotter if seen
equator-on and fainter if seen pole-on, \citeauthor{2005A&A...440..305F} \citeyear{2005A&A...440..305F}).
However, while non-zero overshoot would make the star evolve more rapidly (the star will be younger
at the {\sc TAMS} position), turbulent diffusion due to rapid rotation would make the star evolve 
more slowly (the star would be older at the {\sc TAMS} position, \citeauthor{2000A&A...361..101M} \citeyear{2000A&A...361..101M}),
shifting the observed location in the right direction.
Another possibility which seems more plausible, is a change in chemical composition - in the sense of a higher metallicity.
Interestingly, the disentangled spectra are going in the same sense. This, however, needs to be confirmed by a detailed chemical analysis
of the component spectra.

Because of its location on the tip of the turnoff region, precisely \theto~A was critical to the choice
of the chemical composition suitable for the Hyades cluster in LE01's work. It might be necessary
to look into a few other solutions again. In particular, the higher metallicity and higher helium 
abundance of the ([Fe/H] = +0.19 and Y = 0.27) model, which fitted well the remainder of main-sequence 
stars in the cluster, should be re-investigated. However, such a detailed confrontation is beyond the 
scope of the present work. Furthermore, and for the first time, the availability of the disentangled 
component spectra opens the way to an abundance determination in both components as if they were single 
stars. The resulting disentangled component spectra thus provide two additional ``calibrator'' stars 
useful in a consistent analysis of the chemical composition of cluster members. 

\section{Summary}

The fainter component of the system rotates rapidly such that its lines
are continuously blended with those of the primary during the whole orbit.
This resulted in a broad range of mass ratios published in the literature, ranging
from 0.73 to 0.873. The spectra disentangling technique has proven, under these 
difficult conditions, to be adequate to obtain reliable orbital parameters
as well as the intrinsic spectra of both components. 
These spectra, obtained from a pure mathematical technique without any a priori information about
the component spectral features, permitted us to derive consistent
and accurate atmospheric properties such as the component effective temperatures
(${\rm T_{eff,A}}$=${\rm T_{eff,B}}$= 7800$\pm$170 K and the respective projected
rotational velocities (\vasini~= 68.4 $\pm$ 1.5 \kps~and \vbsini~= 113 $\pm$ 6 \kps).
The value for the mass ratio we found is: q = 0.754 $\pm$ 0.002.
Furthermore, the component spectra were shown to closely reproduce the characteristics 
of an observed reference spectrum (using the single star HD~2628). From the comparison 
with synthetic spectra, we found that enhanced metallicity was needed to model 
\theto~A and B, as expected for two members of the Hyades cluster.

The combined (astrometric--spectroscopic) analysis permitted us to determine the 
orbital parallax (20.90 $\pm$ 0.14 mas) as well as the component masses ($M_A$= 
2.86 $\pm$ 0.06 M$_{\odot}$ and $M_B$= 2.16 $\pm$ 0.02 M$_{\odot}$) with a high accuracy. 
Adopting this new parallax, we obtained the component absolute magnitudes M{\sc v}$_{A}$ 
= 0.33 $\pm$ 0.03 mag and M{\sc v}$_{B}$ = 1.44 $\pm$ 0.03 mag. A further improvement of the 
accuracy on the orbital parameters might be gained if the interval around periastron passage 
were more intensively covered. Indeed, at present, 76\% of the spectra cover half the period 
around apastron, whereas only 24\% cover half the period around periastron. It might also 
be worthwhile to extend the spectral range used for the application of spectra disentangling 
by one order of magnitude (with respect to the present range of 26 \AA) provided that sufficient 
spectra obtained near the epoch of periastron passage are included as well. Obviously, covering this 
crucial orbital phase with more {\sc Elodie} or {\sc Hermes} spectra should lead to a significant
improvement. 
Ideally, the distribution of the spectra should be homogeneous over the entire range of 
Doppler velocities.

{From the confrontation 
between the observed masses and luminosities and evolutionary tracks adopting the 
Hyades composition derived by LE01, we conclude that component B is on the main sequence. 
The interpretation for component~A, however, remains problematic. 
We investigated a number of possibilities (such as convective core overshooting, fast rotation)
to explain the observed discrepancy but it would appear from this first analysis that a 
change in chemical composition - in the sense of a higher metallicity - might be necessary. 

The resulting disentangled component spectra open the way to an abundance analysis in
the components of \theto~ as if they were single stars. This will be useful in the discussion 
concerning the true chemical composition of the Hyades cluster. We intend to perform such 
an analysis making use of the semi-automatic method that we developed for A and F-type stars 
\citep{2009MNRAS.396.1689H}.}

\vspace{0.5cm}
{\small 
{\it Acknowledgements:} 
K.B.V. Torres, Y. Fr\'emat and P. Lampens gratefully acknowledge support
in the framework of the following projects financed by the Belgian Federal Science Policy: the project
"Disentangled Components of Multiple Stars as Laboratories of Stellar Evolution" (Supplementary
Researcher 2008-2009) and the Action-1 project "Pulsation, Chemical Composition and Multiplicity
in Main-Sequence A-F Stars" (Ref. MO/33/018).  
In addition, K.B.V. Torres acknowledges funding from
the Brazilian agencies CNPq, CAPES and FAPEMIG. 
We thank Drs. D. Pourbaix and S. Iliji{\'c} for
supplying their respective codes {\sc FDBinary} and {\sc vbsb2}.
We furthermore thank the Optical Infrared Co-ordination Network
({\sc OPTICON}, a major international collaboration supported by the Research
Infrastructures Programme of the European Commission's Sixth Framework Programme) for the observing
runs being funded by its Trans-National Access Programme (Ref. 2005/042 and 2006/042) as well as Dr. G.
Torres for providing us the spectra collected at the Oak Ridge Observatory.
The Hermes project is a collaboration between the KULeuven, the Universit\'e 
Libre de Bruxelles and the Royal Observatory of Belgium with contributions 
from the Observatoire de Gen\`{e}ve (Switzerland) and the Th\"{u}ringer 
Landessternwarte Tautenburg (Germany). Hermes is funded by the Fund for 
Scientific Research of Flanders (FWO) under the grant G.0472.04, from the 
Research Council of K.U.Leuven under grant GST-B4443, from the Fonds National 
de la Recherche Scientifique under contracts IISN4.4506.05 and FRFC 2.4533.09, 
and financial support from Lotto (2004) assigned to the Royal Observatory of 
Belgium.
The Ond\v{r}ejov Observatory is supported by the project with reference AV0Z10030501.
We thank P. {\v{S}koda}'s collaborators, from Ond\v{r}ejov Observatory, Czech Republic, for
kindly taking 16 spectra of \theto. 
Last but not least, we thank the referee for some very useful suggestions.
This work made use of the 
Simbad astronomical data base operated by the CDS, Strasbourg (France) and the ADS bibliography.

\bibliographystyle{aa}
\bibliography{15166bib}

\Online
\onecolumn 
\tiny{
\longtab{4}{
\tabcolsep 2.5pt
\begin{longtable}{ccr@{}r@{}rcr@{}r@{}rcrccr@{}r@{}rcr@{}r@{}rcr}
\caption{\label{tab:rv:aspall} \theto~AB radial velocities.}\\
\hline\hline\noalign{\smallskip}
HJD $+$       & Orbital & \multicolumn{3}{c}{${\rm RV}_A$}& { (C $-$} & \multicolumn{3}{c}{${\rm RV}_B$} & { (C $-$}& Set & HJD $+$   &Orbital& \multicolumn{3}{c}{${\rm RV}_A$} & { (C $-$} & \multicolumn{3}{c}{${\rm RV}_B$} & { (C $-$} & Set \\ 
  2400000     & phase   & \multicolumn{3}{c}{(\kps)}      &  { FDB)}      & \multicolumn{3}{c}{(\kps)}       &   { FDB)}     &     &   2400000 &phase  & \multicolumn{3}{c}{(\kps)}       & { FDB)}        & \multicolumn{3}{c}{(\kps)}       &  { FDB)}       &\\  
\hline
\endfirsthead
\caption{continued.}\\
\hline\hline\noalign{\smallskip}
HJD $+$       & Orbital & \multicolumn{3}{c}{${\rm RV}_A$}& { (C $-$} & \multicolumn{3}{c}{${\rm RV}_B$} & { (C $-$}& Set & HJD $+$   &Orbital& \multicolumn{3}{c}{${\rm RV}_A$} & { (C $-$} & \multicolumn{3}{c}{${\rm RV}_B$} & { (C $-$} & Set \\ 
  2400000     & phase   & \multicolumn{3}{c}{(\kps)}      &  { FDB)}      & \multicolumn{3}{c}{(\kps)}       &   { FDB)}     &     &   2400000 &phase  & \multicolumn{3}{c}{(\kps)}       & { FDB)}        & \multicolumn{3}{c}{(\kps)}       &  { FDB)}       &\\  
\hline
\endhead
\hline \\ [-6pt]
47844.67396 & 0.68250& { $+$3.217  }& $\pm$ &  1.208&  $+$0.034 & { $-$4.264   }& $\pm$ &   5.456&  $-$0.038 & 1&  50138.48966 & 0.98189& { $+$46.222   }  & $\pm$ &   1.407	&  $-$0.466  & { $-$61.267  }& $\pm$ &  8.022&  $+$0.622 & 1  \\ 
47942.61109 & 0.37842& { $-$9.362  }& $\pm$ &  1.130&  $+$0.022 & { $+$12.409  }& $\pm$ &   5.235&  $-$0.021 & 1&  50140.56345 & 0.99662& { $+$36.590   }  & $\pm$ &   0.605	&  $+$0.265  & { $-$48.499  }& $\pm$ &  2.979&  $-$0.346 & 1  \\   
47957.58553 & 0.48483& { $-$5.451  }& $\pm$ &  0.747&  $+$0.035 & { $+$7.225   }& $\pm$ &   3.442&  $-$0.038 & 1&  50143.47618 & 0.01732& { $+$2.621    }  & $\pm$ &   0.746	&  $+$0.520  & { $-$3.474   }& $\pm$ &  3.329&  $-$0.681 & 1  \\ 
48142.89616 & 0.80161& { $+$11.265 }& $\pm$ &  0.773&  $+$0.001 & { $-$14.933  }& $\pm$ &   3.593&  $+$0.005 & 1&  50146.47843 & 0.03865& { $-$13.453   }  & $\pm$ &   0.677	&  $+$0.093  & { $+$17.831  }& $\pm$ &  3.169&  $-$0.115 & 1  \\ 
48145.81918 & 0.82238& { $+$13.202 }& $\pm$ &  0.701&  $-$0.012 & { $-$17.498  }& $\pm$ &   3.262&  $+$0.023 & 1&  50152.51706 & 0.08156& { $-$19.179   }  & $\pm$ &   0.775	&  $-$0.058  & { $+$25.422  }& $\pm$ &  3.688&  $+$0.086 & 1  \\ 
48158.82114 & 0.91477& { $+$26.806 }& $\pm$ &  0.697&  $-$0.185 & { $-$35.530  }& $\pm$ &   3.274&  $+$0.251 & 1&  50154.54631 & 0.09598& { $-$19.225   }  & $\pm$ &   0.630	&  $-$0.059  & { $+$25.482  }& $\pm$ &  2.996&  $+$0.087 & 1   \\
48162.77544 & 0.94286& { $+$34.391 }& $\pm$ &  0.730&  $-$0.341 & { $-$45.585  }& $\pm$ &   3.533&  $+$0.457 & 1&  53431.26650 & 0.37968& { $-$9.296    }  & $\pm$ &   0.121	&  $+$0.042  & { $+$12.321  }& $\pm$ &  0.559&  $-$0.048 & 2  \\ 
48168.69863 & 0.98495& { $+$45.848 }& $\pm$ &  0.747&  $-$0.343 & { $-$60.771  }& $\pm$ &   4.261&  $+$0.459 & 1&  53431.27540 & 0.37975& { $-$9.293    }  & $\pm$ &   0.121	&  $+$0.042  & { $+$12.318  }& $\pm$ &  0.559&  $-$0.047 & 2  \\ 
48189.80317 & 0.13492& { $-$18.328 }& $\pm$ &  0.557&  $-$0.051 & { $+$24.293  }& $\pm$ &   2.647&  $+$0.076 & 1&  53642.66560 & 0.88184& { $+$20.691   }  & $\pm$ &   0.097	&  $-$0.001  & { $-$27.425  }& $\pm$ &  0.458&  $+$0.008 & 2  \\ 
48191.74188 & 0.14869& { $-$17.862 }& $\pm$ &  0.557&  $-$0.045 & { $+$23.676  }& $\pm$ &   2.646&  $+$0.069 & 1&  53657.52560 & 0.98743& { $+$44.782   }  & $\pm$ &   0.257	&  $-$0.463  & { $-$59.357  }& $\pm$ &  1.464&  $+$0.618 & 2  \\ 
48193.75274 & 0.16298& { $-$17.347 }& $\pm$ &  0.557&  $-$0.039 & { $+$22.994  }& $\pm$ &   2.644&  $+$0.061 & 1&  53667.67700 & 0.05957& { $-$18.027   }  & $\pm$ &   0.097	&  $-$0.070  & { $+$23.893  }& $\pm$ &  0.460&  $+$0.101 & 2  \\ 
48194.82415 & 0.17060& { $-$17.066 }& $\pm$ &  0.557&  $-$0.036 & { $+$22.620  }& $\pm$ &   2.632&  $+$0.056 & 1&  53686.56880 & 0.19381& { $-$16.168   }  & $\pm$ &   0.097	&  $-$0.005  & { $+$21.431  }& $\pm$ &  0.457&  $+$0.016 & 2  \\ 
48198.64843 & 0.19777& { $-$16.039 }& $\pm$ &  0.605&  $-$0.025 & { $+$21.259  }& $\pm$ &   2.855&  $+$0.041 & 1&  53690.58560 & 0.22235& { $-$15.083   }  & $\pm$ &   0.097	&  $+$0.005  & { $+$19.992  }& $\pm$ &  0.454&  $+$0.002 & 2  \\ 
48199.72259 & 0.20540& { $-$15.749 }& $\pm$ &  0.605&  $-$0.022 & { $+$20.874  }& $\pm$ &   2.855&  $+$0.037 & 1&  53711.46230 & 0.37070& { $-$9.619    }  & $\pm$ &   0.145	&  $+$0.042  & { $+$12.751  }& $\pm$ &  0.674&  $-$0.047 & 2  \\ 
48204.81905 & 0.24162& { $-$14.376 }& $\pm$ &  0.726&  $-$0.010 & { $+$19.055  }& $\pm$ &   3.395&  $+$0.021 & 1&  53724.36150 & 0.46236& { $-$6.272    }  & $\pm$ &   0.096	&  $+$0.055  & { $+$8.314   }& $\pm$ &  0.445&  $-$0.064 & 2  \\ 
48206.74581 & 0.25531& { $-$13.862 }& $\pm$ &  0.605&  $-$0.005 & { $+$18.374  }& $\pm$ &   2.829&  $+$0.015 & 1&  53728.37360 & 0.49087& { $-$5.199    }  & $\pm$ &   0.072	&  $+$0.057  & { $+$6.891   }& $\pm$ &  0.333&  $-$0.068 & 2  \\ 
48221.70076 & 0.36158& { $-$9.970  }& $\pm$ &  0.532&  $+$0.019 & { $+$13.215  }& $\pm$ &   2.474&  $-$0.017 & 1&  53739.43640 & 0.56948& { $-$2.069    }  & $\pm$ &   0.120	&  $+$0.063  & { $+$2.743   }& $\pm$ &  0.537&  $-$0.076 & 2  \\ 
48225.71561 & 0.39010& { $-$8.939  }& $\pm$ &  0.531&  $+$0.024 & { $+$11.847  }& $\pm$ &   2.461&  $-$0.024 & 1&  53747.35580 & 0.62575& { $+$0.414    }  & $\pm$ &   0.240	&  $+$0.066  & { $-$0.549   }& $\pm$ &  1.044&  $-$0.080 & 2  \\ 
48226.67264 & 0.39690& { $-$8.692  }& $\pm$ &  0.869&  $+$0.025 & { $+$11.521  }& $\pm$ &   4.016&  $-$0.025 & 1&  53749.39120 & 0.64021& { $+$1.099    }  & $\pm$ &   0.120	&  $+$0.067  & { $-$1.457   }& $\pm$ &  0.525&  $-$0.081 & 2  \\ 
48227.69380 & 0.40416& { $-$8.429  }& $\pm$ &  0.531&  $+$0.027 & { $+$11.172  }& $\pm$ &   2.455&  $-$0.028 & 1&  53759.35790 & 0.71103& { $+$4.857    }  & $\pm$ &   0.096	&  $+$0.065  & { $-$6.438   }& $\pm$ &  0.437&  $-$0.079 & 2  \\ 
48230.72959 & 0.42573& { $-$7.644  }& $\pm$ &  0.530&  $+$0.029 & { $+$10.132  }& $\pm$ &   2.449&  $-$0.030 & 1&  53777.34740 & 0.83886& { $+$14.995   }  & $\pm$ &   0.121	&  $+$0.038  & { $-$19.876  }& $\pm$ &  0.565&  $-$0.044 & 2  \\ 
48232.73228 & 0.43996& { $-$7.122  }& $\pm$ &  0.530&  $+$0.031 & { $+$9.440   }& $\pm$ &   2.447&  $-$0.033 & 1&  53787.30460 & 0.90962& { $+$25.804   }  & $\pm$ &   0.123	&  $-$0.040  & { $-$34.202  }& $\pm$ &  0.584&  $+$0.059 & 2  \\ 
48234.69604 & 0.45392& { $-$6.607  }& $\pm$ &  0.530&  $+$0.032 & { $+$8.757   }& $\pm$ &   2.446&  $-$0.035 & 1&  53793.32650 & 0.95241& { $+$37.826   }  & $\pm$ &   0.101	&  $-$0.221  & { $-$50.137  }& $\pm$ &  0.510&  $+$0.298 & 2  \\ 
48252.69842 & 0.58184& { $-$1.570  }& $\pm$ &  0.529&  $+$0.041 & { $+$2.080   }& $\pm$ &   2.343&  $-$0.047 & 1&  53801.33250 & 0.00930& { $+$14.200   }  & $\pm$ &   0.145	&  $+$0.041  & { $-$18.821  }& $\pm$ &  0.676&  $-$0.048 & 2  \\ 
48257.64582 & 0.61699& { $-$0.016  }& $\pm$ &  0.529&  $+$0.041 & { $+$0.022   }& $\pm$ &   2.300&  $-$0.046 & 1&  53801.34800 & 0.00941& { $+$14.010   }  & $\pm$ &   0.145	&  $+$0.031  & { $-$18.570  }& $\pm$ &  0.676&  $-$0.035 & 2  \\ 
48258.58654 & 0.62368& { $+$0.291  }& $\pm$ &  0.601&  $+$0.040 & { $-$0.386   }& $\pm$ &   2.609&  $-$0.045 & 1&  53801.36060 & 0.00950& { $+$13.857   }  & $\pm$ &   0.169	&  $+$0.057  & { $-$18.367  }& $\pm$ &  0.789&  $-$0.069 & 2  \\ 
48278.54634 & 0.76551& { $+$8.384  }& $\pm$ &  1.415&  $+$0.020 & { $-$11.112  }& $\pm$ &   6.565&  $-$0.019 & 1&  53801.37240 & 0.00958& { $+$13.713   }  & $\pm$ &   0.169	&  $+$0.047  & { $-$18.177  }& $\pm$ &  0.788&  $-$0.056 & 2  \\ 
48279.54381 & 0.77260& { $+$8.910  }& $\pm$ &  0.700&  $+$0.018 & { $-$11.809  }& $\pm$ &   3.249&  $-$0.016 & 1&  53801.38420 & 0.00966& { $+$13.571   }  & $\pm$ &   0.145	&  $+$0.038  & { $-$17.987  }& $\pm$ &  0.676&  $-$0.043 & 2  \\ 
48289.66074 & 0.84449& { $+$15.569 }& $\pm$ &  1.134&  $-$0.029 & { $-$20.636  }& $\pm$ &   5.319&  $+$0.045 & 1&  53802.30060 & 0.01618& { $+$3.621    }  & $\pm$ &   0.830	&  $+$0.055  & { $-$4.800   }& $\pm$ &  3.755&  $-$0.066 & 2  \\ 
48290.50641 & 0.85049& { $+$16.279 }& $\pm$ &  0.653&  $-$0.037 & { $-$21.576  }& $\pm$ &   3.066&  $+$0.056 & 1&  53802.34250 & 0.01647& { $+$3.228    }  & $\pm$ &   0.265	&  $+$0.072  & { $-$4.278   }& $\pm$ &  1.196&  $-$0.088 & 2  \\ 
48310.51233 & 0.99265& { $+$41.466 }& $\pm$ &  0.561&  $+$0.158 & { $-$54.962  }& $\pm$ &   3.007&  $-$0.204 & 1&  53802.36410 & 0.01663& { $+$3.028    }  & $\pm$ &   0.708	&  $+$0.075  & { $-$4.013   }& $\pm$ &  3.197&  $-$0.092 & 2    \\
48325.50344 & 0.09918& { $-$19.195 }& $\pm$ &  0.654&  $-$0.061 & { $+$25.443  }& $\pm$ &   3.111&  $+$0.090 & 1&  53803.27240 & 0.02308& { $-$4.125    }  & $\pm$ &   0.145	&  $+$0.021  & { $+$5.467   }& $\pm$ &  0.662&  $-$0.020 & 2  \\ 
48344.51304 & 0.23425& { $-$14.654 }& $\pm$ &  1.132&  $-$0.011 & { $+$19.423  }& $\pm$ &   5.296&  $+$0.023 & 1&  53803.28070 & 0.02314& { $-$4.179    }  & $\pm$ &   0.120	&  $+$0.038  & { $+$5.539   }& $\pm$ &  0.552&  $-$0.043 & 2  \\ 
48350.50753 & 0.27685& { $-$13.060 }& $\pm$ &  0.629&  $+$0.001 & { $+$17.311  }& $\pm$ &   2.940&  $+$0.007 & 1&  53803.29010 & 0.02321& { $-$4.241    }  & $\pm$ &   0.120	&  $+$0.024  & { $+$5.621   }& $\pm$ &  0.552&  $-$0.024 & 2  \\ 
48992.52662 & 0.83891& { $+$14.945 }& $\pm$ &  0.604&  $-$0.018 & { $-$19.809  }& $\pm$ &   2.827&  $+$0.030 & 1&  53803.29870 & 0.02327& { $-$4.297    }  & $\pm$ &   0.349	&  $+$0.015  & { $+$5.695   }& $\pm$ &  1.599&  $-$0.013 & 2  \\ 
49004.66293 & 0.92514& { $+$29.323 }& $\pm$ &  0.628&  $-$0.212 & { $-$38.867  }& $\pm$ &   2.953&  $+$0.286 & 1&  53803.30710 & 0.02333& { $-$4.351    }  & $\pm$ &   0.145	&  $+$0.032  & { $+$5.768   }& $\pm$ &  0.662&  $-$0.034 & 2  \\ 
49050.57611 & 0.25139& { $-$14.006 }& $\pm$ &  0.895&  $-$0.003 & { $+$18.564  }& $\pm$ &   4.186&  $+$0.012 & 1&  53803.31560 & 0.02339& { $-$4.406    }  & $\pm$ &   0.145	&  $+$0.024  & { $+$5.840   }& $\pm$ &  0.664&  $-$0.024 & 2  \\ 
49267.89835 & 0.79564& { $+$10.761 }& $\pm$ &  0.917&  $+$0.015 & { $-$14.264  }& $\pm$ &   4.262&  $-$0.013 & 1&  53803.32400 & 0.02345& { $-$4.461    }  & $\pm$ &   0.120	&  $+$0.016  & { $+$5.912   }& $\pm$ &  0.554&  $-$0.014 & 2  \\ 
49286.83598 & 0.93021& { $+$30.671 }& $\pm$ &  0.905&  $-$0.228 & { $-$40.655  }& $\pm$ &   4.269&  $+$0.307 & 1&  53803.33260 & 0.02351& { $-$4.516    }  & $\pm$ &   0.265	&  $+$0.031  & { $+$5.986   }& $\pm$ &  1.220&  $-$0.033 & 2  \\ 
49313.76324 & 0.12155& { $-$18.728 }& $\pm$ &  0.920&  $-$0.053 & { $+$24.823  }& $\pm$ &   4.375&  $+$0.079 & 1&  53803.34150 & 0.02357& { $-$4.572    }  & $\pm$ &   0.651	&  $+$0.021  & { $+$6.061   }& $\pm$ &  2.997&  $-$0.019 & 2  \\ 
49329.74569 & 0.23512& { $-$14.617 }& $\pm$ &  0.943&  $-$0.007 & { $+$19.375  }& $\pm$ &   4.414&  $+$0.018 & 1&  53806.29510 & 0.04456& { $-$15.442   }  & $\pm$ &   0.194	&  $-$0.056  & { $+$20.468  }& $\pm$ &  0.913&  $+$0.083 & 2  \\ 
49345.71675 & 0.34860& { $-$10.435 }& $\pm$ &  0.894&  $+$0.021 & { $+$13.831  }& $\pm$ &   4.165&  $-$0.020 & 1&  53806.30400 & 0.04462& { $-$15.458   }  & $\pm$ &   0.194	&  $-$0.059  & { $+$20.489  }& $\pm$ &  0.913&  $+$0.087 & 2  \\ 
49358.67549 & 0.44069& { $-$7.092  }& $\pm$ &  1.075&  $+$0.035 & { $+$9.399   }& $\pm$ &   4.960&  $-$0.039 & 1&  53806.31280 & 0.04469& { $-$15.474   }  & $\pm$ &   0.194	&  $-$0.055  & { $+$20.511  }& $\pm$ &  0.913&  $+$0.082 & 2  \\ 
49371.67152 & 0.53303& { $-$3.575  }& $\pm$ &  1.011&  $+$0.044 & { $+$4.740   }& $\pm$ &   4.575&  $-$0.050 & 1&  53806.32120 & 0.04475& { $-$15.489   }  & $\pm$ &   0.218	&  $-$0.057  & { $+$20.532  }& $\pm$ &  1.028&  $+$0.085 & 2  \\ 
49379.58646 & 0.58927& { $-$1.244  }& $\pm$ &  0.842&  $+$0.046 & { $+$1.649   }& $\pm$ &   3.713&  $-$0.053 & 1&  53806.32980 & 0.04481& { $-$15.506   }  & $\pm$ &   0.218	&  $-$0.060  & { $+$20.552  }& $\pm$ &  1.028&  $+$0.088 & 2  \\ 
49389.64176 & 0.66073& { $+$2.090  }& $\pm$ &  0.866&  $+$0.045 & { $-$2.770   }& $\pm$ &   3.819&  $-$0.052 & 1&  53806.33840 & 0.04487& { $-$15.522   }  & $\pm$ &   0.266	&  $-$0.056  & { $+$20.573  }& $\pm$ &  1.256&  $+$0.082 & 2  \\ 
49404.46549 & 0.76606& { $+$8.432  }& $\pm$ &  0.845&  $+$0.027 & { $-$11.177  }& $\pm$ &   3.920&  $-$0.029 & 1&  53806.34690 & 0.04493& { $-$15.537   }  & $\pm$ &   0.266	&  $-$0.058  & { $+$20.593  }& $\pm$ &  1.256&  $+$0.085 & 2  \\ 
49979.79433 & 0.85423& { $+$16.758 }& $\pm$ &  0.484&  $-$0.021 & { $-$22.212  }& $\pm$ &   2.272&  $+$0.035 & 1&  53806.35260 & 0.04497& { $-$15.547   }  & $\pm$ &   0.691	&  $-$0.055  & { $+$20.608  }& $\pm$ &  3.262&  $+$0.082 & 2  \\ 
49990.88998 & 0.93307& { $+$31.490 }& $\pm$ &  0.503&  $-$0.227 & { $-$41.740  }& $\pm$ &   2.380&  $+$0.306 & 1&  53806.37650 & 0.04514& { $-$15.590   }  & $\pm$ &   0.378	&  $-$0.059  & { $+$20.665  }& $\pm$ &  1.784&  $+$0.087 & 2  \\ 
49992.84490 & 0.94696& { $+$35.800 }& $\pm$ &  0.579&  $-$0.317 & { $-$47.453  }& $\pm$ &   2.839&  $+$0.425 & 1&  53806.38500 & 0.04520& { $-$15.605   }  & $\pm$ &   0.266	&  $-$0.061  & { $+$20.684  }& $\pm$ &  1.256&  $+$0.089 & 2  \\ 
49993.92196 & 0.95462& { $+$38.478 }& $\pm$ &  0.581&  $-$0.382 & { $-$51.002  }& $\pm$ &   3.017&  $+$0.511 & 1&  54905.41312 & 0.85467& { $+$16.850   }  & $\pm$ &   0.069	&  $+$0.017  & { $-$22.340  }& $\pm$ &  0.388&  $-$0.021 & 3  \\ 
49999.87944 & 0.99695& { $+$36.151 }& $\pm$ &  0.680&  $+$0.307 & { $-$47.917  }& $\pm$ &   3.334&  $-$0.402 & 1&  54905.42068 & 0.85472& { $+$16.860   }  & $\pm$ &   0.062	&  $+$0.020  & { $-$22.350  }& $\pm$ &  0.348&  $-$0.023 & 3  \\ 
50000.83846 & 0.00376& { $+$24.852 }& $\pm$ &  0.660&  $+$0.636 & { $-$32.941  }& $\pm$ &   3.141&  $-$0.837 & 1&  54905.42824 & 0.85478& { $+$16.870   }  & $\pm$ &   0.078	&  $+$0.020  & { $-$22.360  }& $\pm$ &  0.442&  $-$0.020 & 3  \\ 
50001.80220 & 0.01061& { $+$12.628 }& $\pm$ &  0.700&  $+$0.690 & { $-$16.737  }& $\pm$ &   3.258&  $-$0.907 & 1&  54906.39330 & 0.86163& { $+$17.750   }  & $\pm$ &   0.081	&  $+$0.011  & { $-$23.530  }& $\pm$ &  0.456&  $-$0.011 & 3  \\ 
50002.85421 & 0.01809& { $+$ 1.670 }& $\pm$ &  0.842&  $+$0.517 & { $-$2.214   }& $\pm$ &   3.714&  $-$0.678 & 1&  54906.39728 & 0.86166& { $+$17.760   }  & $\pm$ &   0.078	&  $+$0.018  & { $-$23.540  }& $\pm$ &  0.443&  $-$0.016 & 3  \\ 
50004.84002 & 0.03220& { $-$10.464 }& $\pm$ &  0.701&  $+$0.192 & { $+$13.871  }& $\pm$ &   3.265&  $-$0.246 & 1&  54906.40149 & 0.86169& { $+$17.760   }  & $\pm$ &   0.078	&  $+$0.014  & { $-$23.540  }& $\pm$ &  0.443&  $-$0.012 & 3  \\ 
50006.86013 & 0.04655& { $-$15.840 }& $\pm$ &  0.653&  $+$0.030 & { $+$20.996  }& $\pm$ &   3.083&  $-$0.031 & 1&  55079.70896 & 0.09318& { $-$19.240   }  & $\pm$ &   0.095	&  $-$0.058  & { $+$25.500  }& $\pm$ &  0.537&  $+$0.083 & 3  \\ 
50007.84894 & 0.05358& { $-$17.204 }& $\pm$ &  0.654&  $-$0.007 & { $+$22.803  }& $\pm$ &   3.104&  $+$0.018 & 1&  55079.71082 & 0.09319& { $-$19.240   }  & $\pm$ &   0.095	&  $-$0.058  & { $+$25.500  }& $\pm$ &  0.537&  $+$0.083 & 3  \\ 
50008.83881 & 0.06061& { $-$18.095 }& $\pm$ &  0.581&  $-$0.031 & { $+$23.984  }& $\pm$ &   2.761&  $+$0.050 & 1&  55080.70855 & 0.10028& { $-$19.180   }  & $\pm$ &   0.095	&  $-$0.059  & { $+$25.420  }& $\pm$ &  0.537&  $+$0.085 & 3  \\ 
50009.86739 & 0.06792& { $-$18.678 }& $\pm$ &  0.581&  $-$0.047 & { $+$24.758  }& $\pm$ &   2.762&  $+$0.071 & 1&  55080.71054 & 0.10030& { $-$19.180   }  & $\pm$ &   0.095	&  $-$0.059  & { $+$25.420  }& $\pm$ &  0.537&  $+$0.085 & 3  \\ 
50020.84972 & 0.14596& { $-$17.951 }& $\pm$ &  0.557&  $-$0.040 & { $+$23.794  }& $\pm$ &   2.646&  $+$0.062 & 1&  55080.71256 & 0.10031& { $-$19.170   }  & $\pm$ &   0.119	&  $-$0.050  & { $+$25.420  }& $\pm$ &  0.671&  $+$0.085 & 3  \\ 
50021.84975 & 0.15307& { $-$17.701 }& $\pm$ &  0.581&  $-$0.037 & { $+$23.462  }& $\pm$ &   2.761&  $+$0.058 & 1&  55082.70400 & 0.11446& { $-$18.900   }  & $\pm$ &   0.095	&  $-$0.044  & { $+$25.050  }& $\pm$ &  0.537&  $+$0.066 & 3  \\ 
50026.75460 & 0.18792& { $-$16.406 }& $\pm$ &  0.629&  $-$0.022 & { $+$21.746  }& $\pm$ &   2.970&  $+$0.037 & 1&  55082.70576 & 0.11447& { $-$18.900   }  & $\pm$ &   0.095	&  $-$0.044  & { $+$25.050  }& $\pm$ &  0.537&  $+$0.066 & 3  \\ 
50027.82940 & 0.19556& { $-$16.116 }& $\pm$ &  0.581&  $-$0.018 & { $+$21.361  }& $\pm$ &   2.742&  $+$0.033 & 1&              &        &                     &       &              &        &                  &       &       &           &    \\ 
\hline                                
\end{longtable}                       
\tablefoot{The radial velocities (RV) were derived relatively to the center of mass
($\gamma = 39.3$ \kps) with the code {\sc FDBinary}. (C-FDB) refers
to the differences in the sense {\bf C}ombined astrometric--spectroscopic solution - {\bf FDB}inary orbit.
The uncertainties were computed as explained in Sect.~5.1. 
The Heliocentric Julian Day (HJD) and orbital phase corresponding to each radial velocity are also included.
Set 1: CfA data; Set 2: {\sc Elodie} data; Set 3: {\sc Hermes} data.}
}                                     
}                                     
                                      
\end{document}